\documentclass[twocolumn]{aastex62}
\usepackage{amsmath,amstext}
\usepackage[T1]{fontenc}
\usepackage{apjfonts} 
\usepackage[figure,figure*]{hypcap}
\usepackage{commath}

   \newcommand{\rah}{$^{\mbox{\scriptsize h}}$}
   \newcommand{\ram}{$^{\mbox{\scriptsize m}}$}
   
   \newcommand{\decd}{$^{\circ}$}
   \newcommand{\decm}{$'$}

   \newcommand{\beam}{$\theta_{\mbox{\scriptsize maj}}\times\theta_{\mbox{\scriptsize min}}$}

   \newcommand{\amax}{ $a_{\mbox{\scriptsize max}}$ }
   
   \newcommand{\dustang}{$\Omega_{\mbox{\scriptsize dust}}$}

\shorttitle{au scales model for FU\,Ori}
\shortauthors{Liu, H.-B. et al.}

\begin{document}

%\title{Extra red millimeter spectrum of the inner protoplanetary disk may infer (sub-)hundred micron maximum grain size}
\title{Diagnosing 0.1-10 au scale morphology of the FU\,Ori disk using ALMA and VLTI/GRAVITY}

% ALMA projecdt co-Is:
% Ruobing Dong, Michael Dunham, Roberto Galvan-Madrid, Yasuhiro Hasegawa, Thomas Henning, Agnes Kospal, Luis Rodriguez, Michihiro Takami, Marco Tazzari, Zhaohuan, Jun Hashimoto
% Guillaume Bourdarot <guillaume.bourdarot@univ-grenoble-alpes.fr>
% Jean-Philippe Berger <Jean-Philippe.Berger@univ-grenoble-alpes.fr>

\correspondingauthor{Hauyu Baobab Liu}
\email{hyliu@asiaa.sinica.edu.tw}

\author[0000-0003-2300-2626]{Hauyu Baobab Liu}
\affiliation{Academia Sinica Institute of Astronomy and Astrophysics, P.O. Box 23-141, Taipei 10617, Taiwan} 

\author[0000-0003-2125-0183]{Antoine M\'erand}
\affiliation{European Southern Observatory (ESO), Karl-Schwarzschild-Str. 2, 85748, Garching, Germany}

\author[0000-0003-1665-5709]{Joel D. Green}
\affiliation{Space Telescope Science Institute, Baltimore, MD 21218, USA ; Department of Astronomy, The University of Texas at Austin, Austin, TX 78712, USA}

\author[0000-0003-2953-755X]{Sebasti\'an P\'erez}
\affiliation{Universidad de Santiago de Chile, Av. Libertador Bernardo O'Higgins 3363, Estaci\'on Central, Santiago, Chile}

\author[0000-0001-5073-2849]{Antonio S. Hales}
\affiliation{Joint ALMA Observatory, Avenida Alonso de Córdova 3107, Vitacura 7630355, Santiago, Chile ; National Radio Astronomy Observatory, 520 Edgemont Road, Charlottesville, VA 22903-2475, USA}

\author[0000-0001-8227-2816]{Yao-Lun Yang}
\affiliation{The University of Texas at Austin, Department of Astronomy, 2515 Speedway, Stop C1400, Austin, TX 78712, USA}

\author[0000-0003-0749-9505]{Michael M. Dunham}
\affiliation{Department of Physics, State University of New York at Fredonia, 280 Central Avenue, Fredonia, NY 14063, USA}

\author{Yasuhiro Hasegawa}
\affiliation{Jet Propulsion Laboratory, California Institute of Technology, Pasadena, CA 91109, USA}

\author[0000-0002-1493-300X]{Thomas Henning}
\affiliation{Max Planck Institute for Astronomy (MPIA), K\"onigstuhl 17, 69117 Heidelberg, Germany}

\author[0000-0003-1480-4643]{Roberto Galv\'an-Madrid}
\affiliation{Instituto de Radioastronom\'ia y Astrof\'isica (IRyA), UNAM, Apdo. Postal 72-3 (Xangari), Morelia, Michoac\'an 58089, Mexico}

\author[0000-0001-7157-6275]{\'Agnes K\'osp\'al}
\affiliation{Konkoly Observatory, Research Centre for Astronomy and Earth Sciences, Hungarian Academy of Sciences, Konkoly-Thege Mikl\'os \'ut 15-17, 1121 Budapest, Hungary}
\affiliation{Max Planck Institute for Astronomy (MPIA), K\"onigstuhl 17, 69117 Heidelberg, Germany}

\author[0000-0001-9248-7546]{Michihiro Takami}
\affiliation{Academia Sinica Institute of Astronomy and Astrophysics, P.O. Box 23-141, Taipei 10617, Taiwan} 

\author[0000-0001-5073-2849]{Eduard I. Vorobyov}
\affiliation{Department of Astrophysics, University of Vienna, Vienna, 1180, Austria}
\affiliation{Research Institute of Physics, Southern Federal University, Rostov-on-Don, 344090 Russia}

\author[0000-0003-3616-6822]{Zhaohuan Zhu}
\affiliation{Department of Physics and Astronomy, University of Nevada, Las Vegas, 4505 S. Maryland Pkwy, Las Vegas, NV 89154, USA}

\begin{abstract}
We report new Atacama Large Millimeter/submillimeter Array Band 3 (86-100 GHz; $\sim$80 mas angular resolution) and Band 4 (146-160 GHz; $\sim$50 mas angular resolution) observations of the dust continuum emission towards the archetypal and ongoing accretion burst young stellar object FU\,Ori, which simultaneously covered its companion, FU\,Ori\,S.
In addition, we present near-infrared (2-2.45 $\mu$m) observations of FU\,Ori taken with the General Relativity Analysis via VLT InTerferometrY (GRAVITY; $\sim$1 mas angular resolution) instrument on the Very Large Telescope Interferometer (VLTI).
We find that the emission in both FU\,Ori and FU\,Ori\,S at (sub)millimeter and near infrared bands is dominated by structures inward of $\sim$10 au radii.
We detected closure phases close to zero from FU\,Ori with VLTI/GRAVITY, which indicate the source is approximately centrally symmetric and therefore is likely viewed nearly face-on.
% However, the structure of the inner few au disk regions may be more complicated than an axisymmetric thin disk in Keplerian rotation.
Our simple model to fit the GRAVITY data shows that the inner 0.4 au radii of the FU\,Ori disk has a triangular spectral shape at 2-2.45 $\mu$m, which is consistent with the H$_{2}$O and CO absorption features in a $\dot{M}\sim$10$^{-4}$ $M_{\odot}\,yr^{-1}$, viscously heated accretion disk.
At larger ($\sim$0.4-10 au) radii, our analysis shows that viscous heating may also explain the observed (sub)millimeter and centimeter spectral energy distribution when we assume a constant, $\sim$10$^{-4}$ $M_{\odot}\,yr^{-1}$ mass inflow rate in this region.
This explains how the inner 0.4 au disk is replenished with mass at a modest rate, such that it neither depletes nor accumulates significant masses over its short dynamic timescale.
Finally, we tentatively detect evidence of vertical dust settling in the inner 10 au of the FU\,Ori disk, but confirmation requires more complete spectral sampling in the centimeter bands.
% Moreover, the observed spectral energy distributions of FU\,Ori between 29-346 GHz may imply that in the inner few au regions dust at the disk mid-plane is hotter than dust at the disk surface.
% Such a vertically inverted dust temperature profile is not required to explain the FU\,Ori\,S data.
% We posit that the inner few au region of FU\,Ori is undergoing (thermal) instabilities: the disk is significantly heated due to viscous dissipation and has a large scale height.
% Close to the disk mid-plane, the circumstellar material may be weakly thermally ionized, and the dust may be partially sublimated.
% This may also make it easier to radiatively heat the disk from close to the mid-plane. 
\end{abstract}

\keywords{stars: individual (FU\,Ori) --- protoplanetary disks}

\section{Introduction}\label{sec:introduction}

Understanding the physical mechanisms of protostellar accretion is fundamentally important in studies of star formation.  
Optical and near infrared surveys have shown that young stellar objects (YSOs) are 10-100 times underluminous with respect to the expected luminosity from steady accretion \citep{Kenyon1995,Evans2009}, which indicates that YSOs may accrete episodically \citep{Dunham2012}.
If episodic accretion is a widespread phenomenon during the YSO phases, it should manifest observationally. This is consistent with the discoveries of two types of YSOs in outburst: the FU Orionis (FUor) and the EX Lupi (EXor) objects, which are characterized by a rapid large increase in their optical and infrared (OIR) brightness \citep[for reviews, see][]{Hartmann1996ARA&A,Herbig2007,Audard2014}. 

FUors have outburst durations of decades to centuries \citep{Hartmann1996ARA&A}.
During the outburst state, their optical brightness can increase by $\sim$4 magnitudes or more.
Models suggest that the accretion rates of these YSOs vary from 10$^{-7}$ M$_{\odot}$ yr$^{-1}$ in the low (T Tauri) accretion state to 10$^{-4}$ M$_{\odot}$ yr$^{-1}$ in the high (FUors) accretion state \citep{Hartmann1996ARA&A}.
While accretion processes in quiescent T Tauri stars are generally understood as magnetospheric streams from the inner disk \citep[e.g.,][]{Koenigl1991ApJ, Calvet2000}, how the gas and dust reservoirs immediately around the FUors are different (or altered) compared with quiescent T Tauri stars is not yet well-understood.
This limits our understanding of the outburst triggering mechanisms and the consequences of them.

To shed light on this issue, we have performed high angular resolution observations towards the archetypal FU Orionis object, FU\,Ori, using the Atacama Large Millimeter Array (ALMA) and the General Relativity Analysis via VLT InTerferometrY (GRAVITY) instrument of the Very Large Telescope Interferometer (VLTI).
Throughout this manuscript, we assume the distance of FU\,Ori to be $d\sim$416 pc, according to the parallax measurement published in the second data release of the Gaia space telescope \citep{Gaia2018A&A}.
According to the prior-assisted parallax distances measurements of \citet{Bailer-Jones2018AJ}, we quote a nominal $\pm$2\% distance uncertainty, which will not qualitatively affect our analysis.
The observations are introduced in Section \ref{sec:obs},
and the results are presented in Section \ref{sec:results}.
By jointly analyzing these new observations with the previous (sub)millimeter observations of the ALMA, the Submillimeter Aray (SMA)\footnote{The Submillimeter Array is a joint project between the Smithsonian
Astrophysical Observatory and the Academia Sinica Institute of Astronomy and Astrophysics, and is funded by the Smithsonian Institution and
the Academia Sinica \citep{Ho2004}} and the NRAO\footnote{The National Radio Astronomy Observatory is a facility of the National Science Foundation operated under cooperative agreement by Associated Universities, Inc.} Karl G. Jansky Very Large Array (JVLA), and the infrared spectra taken with the {\it Spitzer} and {\it Herschel}\footnote{Herschel is an ESA space observatory with science instruments provided by European-led Principal Investigator consortia and with important participation from NASA.} space telescopes,
% we discovered that the inner $\sim$10 AU region of FU\,Ori may have a vertically inverted thermal profile as compared to the normal T Tauri disks,
% which may be related to the instabilities triggered during the accretion outburst.
our interpretation and the further discussion about the physical implications are provided in Section \ref{sec:discussion}.
Our conclusion is given in Section \ref{sec:conclusion}.
We refer to \citet{Berger2012} for a review of the convention and terminology for the optical and infrared interferometry technique. % given that presently more readers may be familiar with the radio interferometry.

% page 146 of ALMA Technical Handbook: The flux density models are determined between 30 and 2000 GHz, and are accurate to about 5% at the lower frequencies. This increases at frequencies above Band 7, however, since the spectral extrapolation of the grid source flux densities is not accurate; in Bands 9 and 10, the conversion from K to Jy is accurate to about 20\%.

\section{Observations}\label{sec:obs}
We introduce the archival {\it Spitzer} and {\it Herschel} spectra and the VLTI/GRAVITY observations in Sections \ref{sub:spitzer} and \ref{sub:gravity_obs}.
We provide detailts of our ALMA observations in Section \ref{sub:alma_obs}.

\subsection{Spitzer and Herschel spectra}\label{sub:spitzer}
% We took the {\it Spitzer}-IRS spectrum of FU\,Orionis from \citet{Green2016ApJ}. 
The {\it Herschel}/PACS and SPIRE spectra were taken from the COPS-DIGIT-FOOSH (CDF) archive, a high-level data product provided to the {\it Herschel} Science Archive (see \citealt{Green2016AJ} for details).
Because the source size at these wavelengths is comparable to the SPIRE beam size, and because of the lack of background subtraction, the spectra of the two modules of SPIRE instruments (SLW and SSW) are often mismatched.  To resolve this discrepancy, \citet{Green2016AJ} apply the Semi-Extended Correction Tool in \textsc{hipe} \citep{Ott2010,Wu2013} to calibrate the SPIRE spectra by modeling the source size.  The best-fit source size, 23\farcs{5} for FU\,Ori, is then convolved with the beam profile of SPIRE, which is a function of wavelength \citep{Mikawa2013}.  
Therefore, the resulting SPIRE spectrum represents the emission from different apertures at given wavelengths, which correspond to the convolved sizes of the beam and the source size \citep{Yang2018}.  
For example, the aperture sizes are 29\farcs{9}, 34\farcs{5}, and 43\farcs{6} at 250~$\mu$m, 350~$\mu$m, and 500~$\mu$m, respectively.
We refer to Section 2.2 of \citet{Green2016AJ} for a complete description of the data reduction.

%In the following sections we describe the technical details of the VLTI/GRAVITY and the ALMA observations.

\subsection{VLTI/GRAVITY observations}\label{sub:gravity_obs}

FU\,Ori was observed by VLTI/GRAVITY \citep{2017A&A...602A..94G} on 2016 November 25 and 26.
These observations were part of the consortium Guaranteed Time Observations (Program ID 098.C-0765). 
The observations were carried out at both medium and high spectral resolution, although only the medium resolution data achieved sufficient signal to noise ratio (S/N) for analysis. 
The medium spectral resolution setting covered the whole near-infrared K-band with a spectral resolution of $\sim$500. 

\begin{figure}
  \vspace{-0.1cm}
  \hspace{-1.1cm}
  \includegraphics[width=11cm, trim={0 0 0 1.2cm},clip]{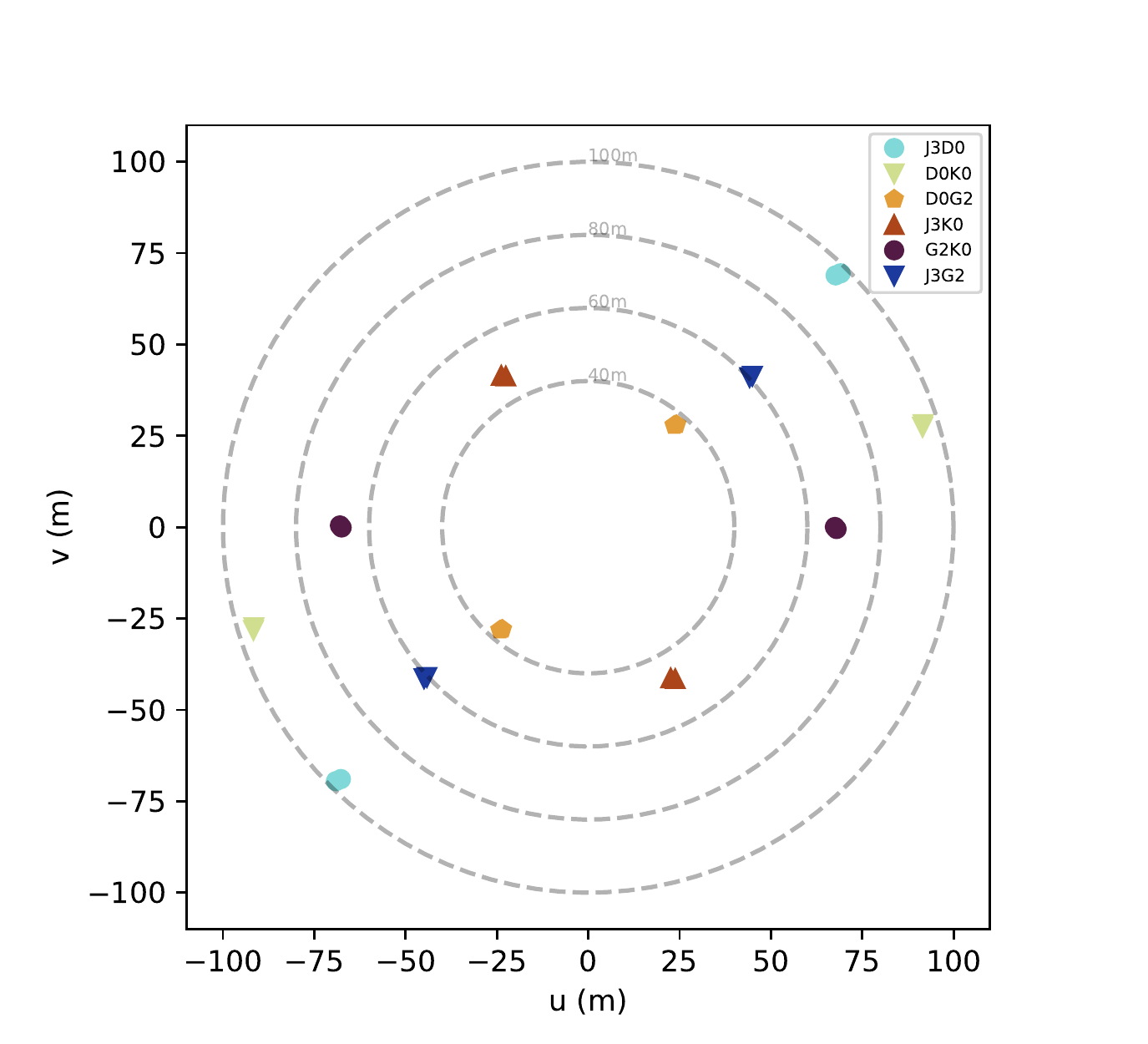}
  \vspace{-0.7cm}
  \caption{Projected baselines ({\it uv} plane) of the VLTI/GRAVITY observations towards FU\,Ori.}
  \label{Fig_uv}
\end{figure}

The telescopes chosen for these observations {\bf were} the {\it medium} Auxiliary Telescopes (AT) configurations.
This configuration includes the stations K0-G2-D0-J3, which led to baselines ranging from $\sim$40 to $\sim$100 meters (Figure \ref{Fig_uv}).

The calibrator star observed concurrently was HD\,38494, which is a K2 star of unknown luminosity class.
Its photometric angular diameter was estimated to be $\theta_\mathrm{UD}=0.71\pm0.06$ mas according to the Jean-Marie Mariotti Center Stellar Diameters Catalogue (JSDC) \citep{2017yCat.2346....0B}.
It is nearly unresolved for our observations. 
The visibility of the fringes is expected to range from 0.971 to 0.996.
The uncertainty of the photometric angular diameter of HD\,38494 leads to a bias in the reduced data of at most 0.005 in visibility.
We reduced the data using the GRAVITY pipeline \citep{2014SPIE.9146E..2DL} version 1.0.11.
We note that the direct observables from VLTI/GRAVITY are normalized to the total flux.
The reduced spectrum does not clearly present emission lines (Figure \ref{Fig_uv}).
% However, CO band-heads are visible as absorption lines.

\subsection{ALMA observations}\label{sub:alma_obs}
We have performed ALMA Bands 3 and 4 observations towards FU\,Ori, which also covered its $\sim$0$\farcs$5 separation companion, FU\,Ori\,S (Project code: 2017.1.00388.S, PI: Hauyu Baobab Liu).
The pointing and phase referencing center was R.A. (J2000) = 05\rah45\ram22$\fs$375, and Decl. (J2000) = +09\decd04\decm12$\farcs$400.
The {\it uv} distance ranges covered by these observations are $\sim$100 m -- 13 km.
The correlators were configured to cover four 1.875 GHz wide spectral windows with a 976.562 kHz channel spacing.

The Band 3 observations were carried out on 2017 November 08.
The four spectral windows were centered on the sky frequencies of 86.000, 87.863, 98.196, and 100.001 GHz.
We observed the quasar J0510+1800 for absolute flux and passband calibrations, and J0547+1223 for complex gain calibrations.

The Band 4 observations were carried out on 2017 November 07.
The four spectral windows were centered on the sky frequencies of 146.001, 147.863, 158.196, and 160.001 GHz.
We observed the quasar J0510+1800 for absolute flux and passband calibrations, and J0536+0944 for complex gain calibrations.

We manually calibrated and phase self-calibrated these data using the CASA software package \citep{McMullin2007} version 5.4.0. 
When performing absolute flux scaling, we assumed that J0510+1800 has a 2.0 Jy absolute flux and a $-$0.30 spectral index at the reference frequency 93.015 GHz; and has a 1.6 Jy absolute flux and a $-$0.4 spectral index at the reference frequency 153.016 GHz. 
These assumptions were based on interpolating the calibrator grid survey measurements.
We produced the Briggs Robust = 0 weighted continuum images from line-free spectral channels using the CASA task {\tt clean}.
For each of the two observed bands, we created images for each of the four spectral windows  separately using the multi-frequency synthesis (MFS) method, setting the parameter {\it nterm}=1.
The four spectral windows in each band achieved comparable root-mean-square (RMS) noise levels and angular resolutions.
At Band 3, the spectral window centered at 100 GHz achieved a \beam=0$''$.082$\times$0$''$.075 (P.A.=-79$^{\circ}$) synthesized beam and a 69 $\mu$Jy\,beam$^{-1}$ RMS noise level; at Band 4, the spectral window centered at 160 GHz achieved a \beam=0$''$.047$\times$0$''$.043 (P.A.=57$^{\circ}$) synthesized beam and a 72 $\mu$Jy\,beam$^{-1}$ RMS noise level.
In each band, the synthesized beam sizes at other spectral windows are inversely proportional to their central frequencies.

\begin{figure*}
  \hspace{-0.6cm}
  \includegraphics[width=19cm]{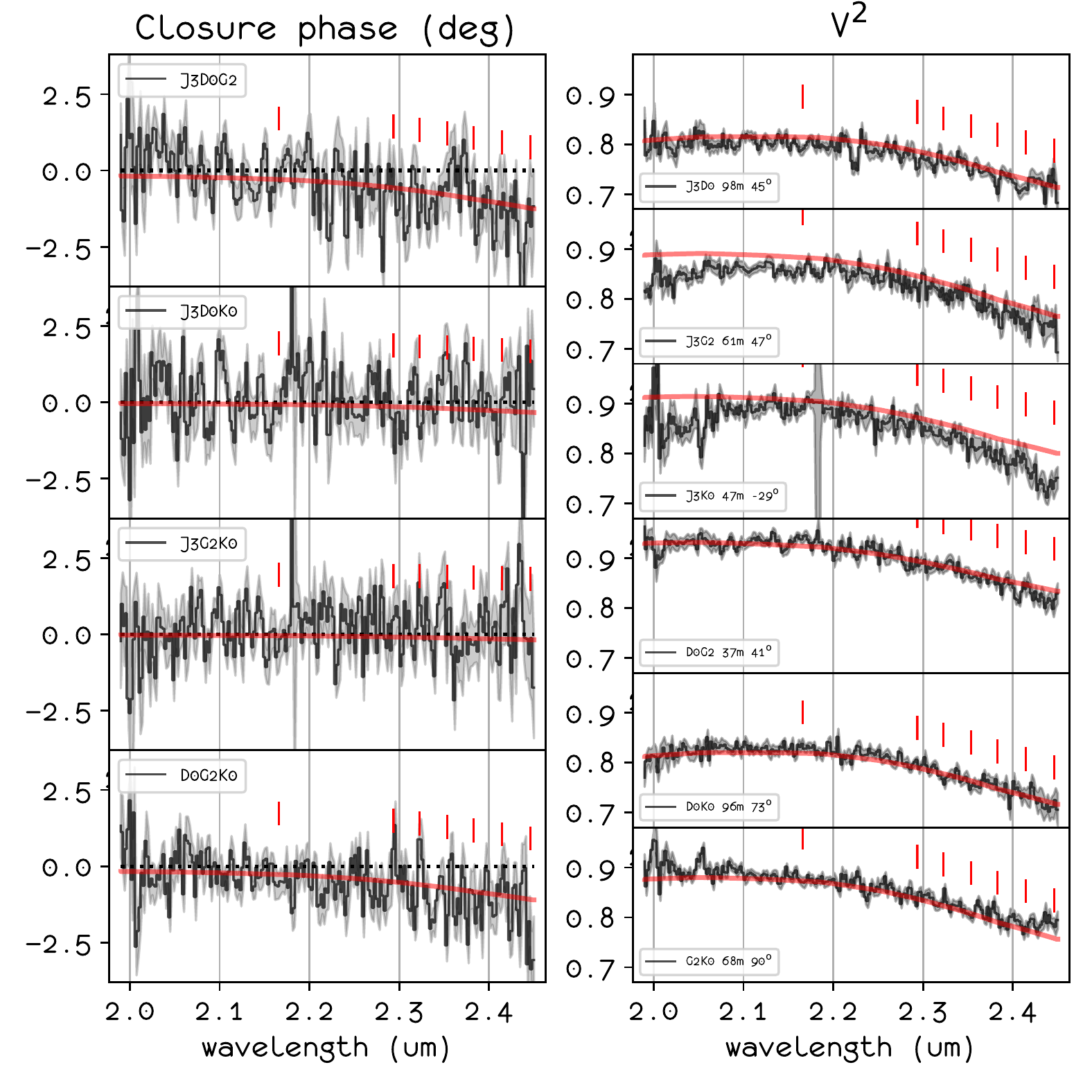}
  \vspace{-0.3cm}
  \caption{Visibility data for FU\,Ori taken with VLTI/GRAVITY. The left and right panels show the closure phase (in degree units) and the squared visibility (normalized to 1) as function of wavelength, respectively. The red curves are our best fit model to the VLTI/GRAVITY data, which is composed of the marginally spatially resolved compact and extended sources (i.e., Model \#4 in Table \ref{tab_models}; see Figure \ref{fig_flux} for more details of the model). 
  For each closure phase, the names of the involved telescopes are labeled, which can be referenced from Figure \ref{Fig_uv}.
  In each panel, the name, length and orientation of the baseline are labeled.
  The vertical red line segments indicate the wavelengths of the Bracket $\gamma$ transition of Hydrogen (2.16612 $\mu$m) and the CO band heads (2.2935, 2.3227, 2.3535, 2.3829, 2.4142, 2.4461 $\mu$m).
  }
  \label{fig_data}
\end{figure*}

\section{Results}\label{sec:results}

\subsection{VLTI/GRAVITY data}\label{sub:gravityresult}

Figure \ref{fig_data} shows the reduced VLTI/GRAVITY data on FU\,Ori.
The closure phases (CP) are smaller than $\pm$2.5 degrees. 
In addition, the overall scatter of the CP is less that 1 degree. 
This indicates that, on the spatial scales resolved by our VLTI/GRAVITY observations, FU\,Ori appears approximately centro-symmetric.
Overall, the squared visibilities have a fairly high level, ranging from 0.7 to 0.9.
The variations of the squared visibilities with wavelength are similar for all baselines, irrespective of baseline lengths and orientations: The squared visibilities are approximately constant from wavelength 2.0 to 2.2 $\mu$m, and then drop by about 0.1 from 2.2 to 2.45$\mu$m.

The fact that the differential visibility variations do not seem to depend on baseline lengths indicates that the intensity distributions may be approximated by a compact component (hereafter VLTI-compact) at the center and a more extended centro-symmetric component (hereafter VLTI-extended)\footnote{Note that the two infrared emission components resolved by the VLTI/GRAVITY observations are both more compact than what were detected by JVLA and ALMA. Our terminology is to distinguish them from the spatially more extended (sub)millimeter and centimeter sources.}.
Including a structure which is nearly resolved out by all baselines leads to the observed $<$1.0 squared visibilities.
In this case, the detected values of the squared visibilities depend mostly on the flux ratios of the VLTI-compact and the VLTI-extended components. 

To give a qualitative sense, if we define VLTI-unresolved as having a visibility higher than 0.99 and VLTI-resolved as having a visibility amplitude less than 0.01, then for our longest, $\sim$100 meter baseline, a  uniform disk with $\lesssim$0.3 mas diameter is VLTI-unresolved.
For our shortest, $\gtrsim$30 meter baseline, a two dimensional Gaussian with $\sim$17 mas full width at half maximum is VLTI-resolved.
The visibility amplitude of a 1 mas compact uniform disk ranges from 0.93 to 0.99 for the baselines ranging from 60~m to 100~m. 
The visibility amplitude of a FWHM$=$8 mas Gaussian ranges from $\sim$0.0001 to $\sim$0.15.

\begin{deluxetable}{ ccccc }

  \tablecaption{Fluxes measurements from ALMA\label{tab:almaflux}}
  \tablehead{
    &  \multicolumn{2}{ c }{FU\,Ori}  &  \multicolumn{2}{ c }{FU\,Ori\,S} 
  }
  \startdata
            Frequency & Flux & Flux error & Flux & Flux error\\
            (GHz)     & (mJy) & (mJy) & (mJy) & (mJy) \\
            \noalign{\smallskip}
            \hline
            \multicolumn{5}{ c }{( Band 3 )} \\
            86.001  & 2.1  & 0.088  & 1.2 &  0.13 \\
            87.863  & 2.3  & 0.13   & 1.4 &  0.12 \\
            98.196  & 2.8  & 0.11   & 1.5 &  0.11 \\
            100.001  & 3.0  & 0.11  & 1.6 &  0.14 \\\hline
            \multicolumn{5}{ c }{( Band 4 )} \\
            146.002  & 6.2  & 0.18  & 3.5 &  0.15 \\
            147.863  & 6.2  & 0.17  & 3.5 &  0.18 \\
            158.196  & 7.0  & 0.17  & 4.0 &  0.19 \\
            160.002  & 7.4  & 0.20  & 4.0 &  0.19 \\
  \enddata
\end{deluxetable}

\begin{deluxetable}{ p{4cm} p{1.8cm} p{1.8cm} }

  \tablecaption{(Sub)millimeter spectral indices\label{tab:spectralindex}}
  \tablehead{
    &  FU\,Ori  &  FU\,Ori\,S
  }
  \startdata
            Frequency range (GHz) &  \multicolumn{2}{ c }{Spectral index ($\alpha$)} \\
            \noalign{\smallskip}
            \hline
            29-37 GHz   & 1.6$\pm$0.4  & 1.4$\pm$0.4 \\
            29-100 GHz  & 2.5$\pm$0.05  & 2.7$\pm$0.05 \\
            86-160 GHz  & 2.0$\pm$0.07  & 1.9$\pm$0.07 \\
            146-232 GHz & 2.0$\pm$0.05  & 2.0$\pm$0.05 \\
            218-346 GHz & 2.9$\pm$0.2  & 2.2$\pm$0.2 \\
  \enddata
\end{deluxetable}

\begin{figure*}
    \hspace{-0.8cm}
    \begin{tabular}{ p{9cm} p{9cm} }
      \includegraphics[width=9cm]{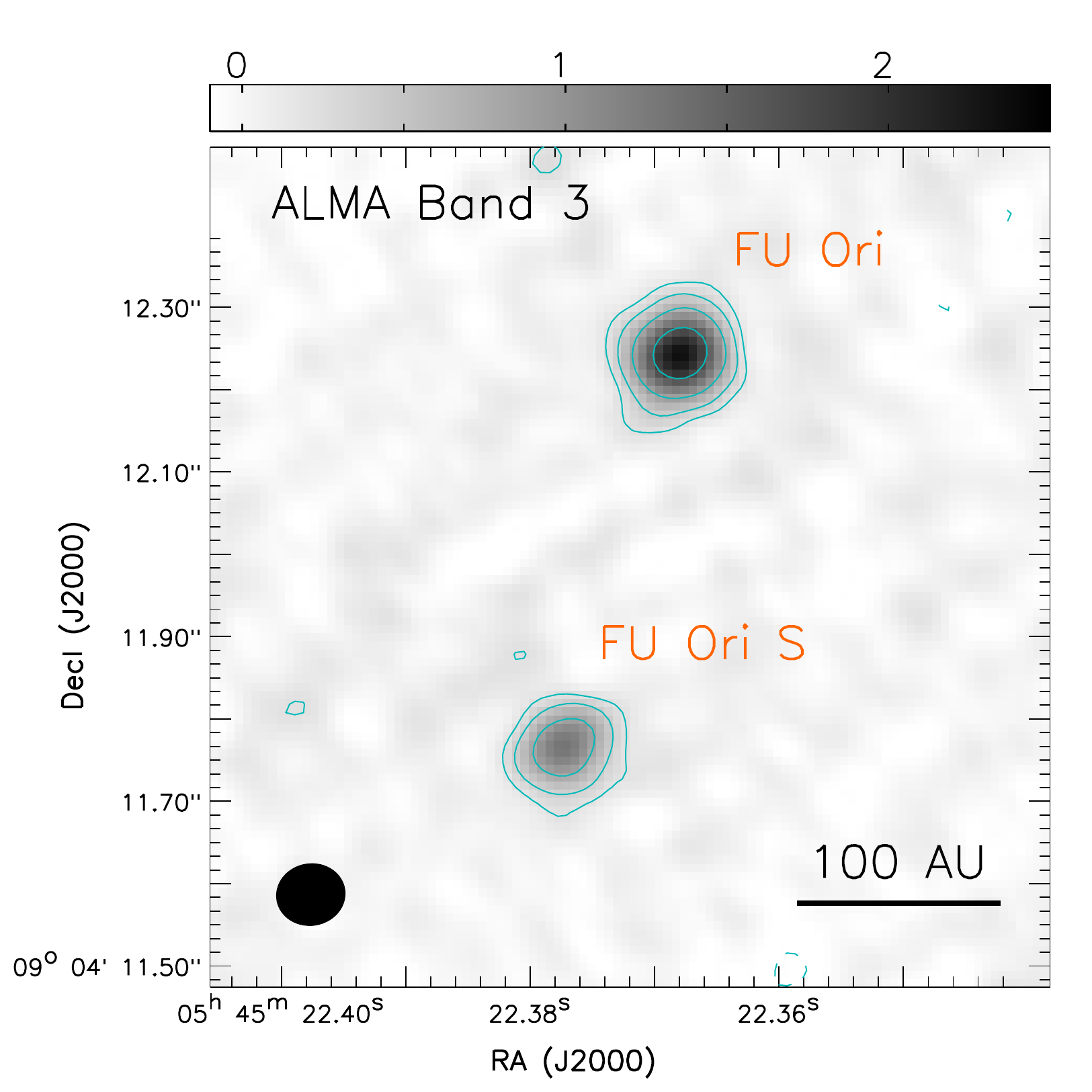} &
      \includegraphics[width=9cm]{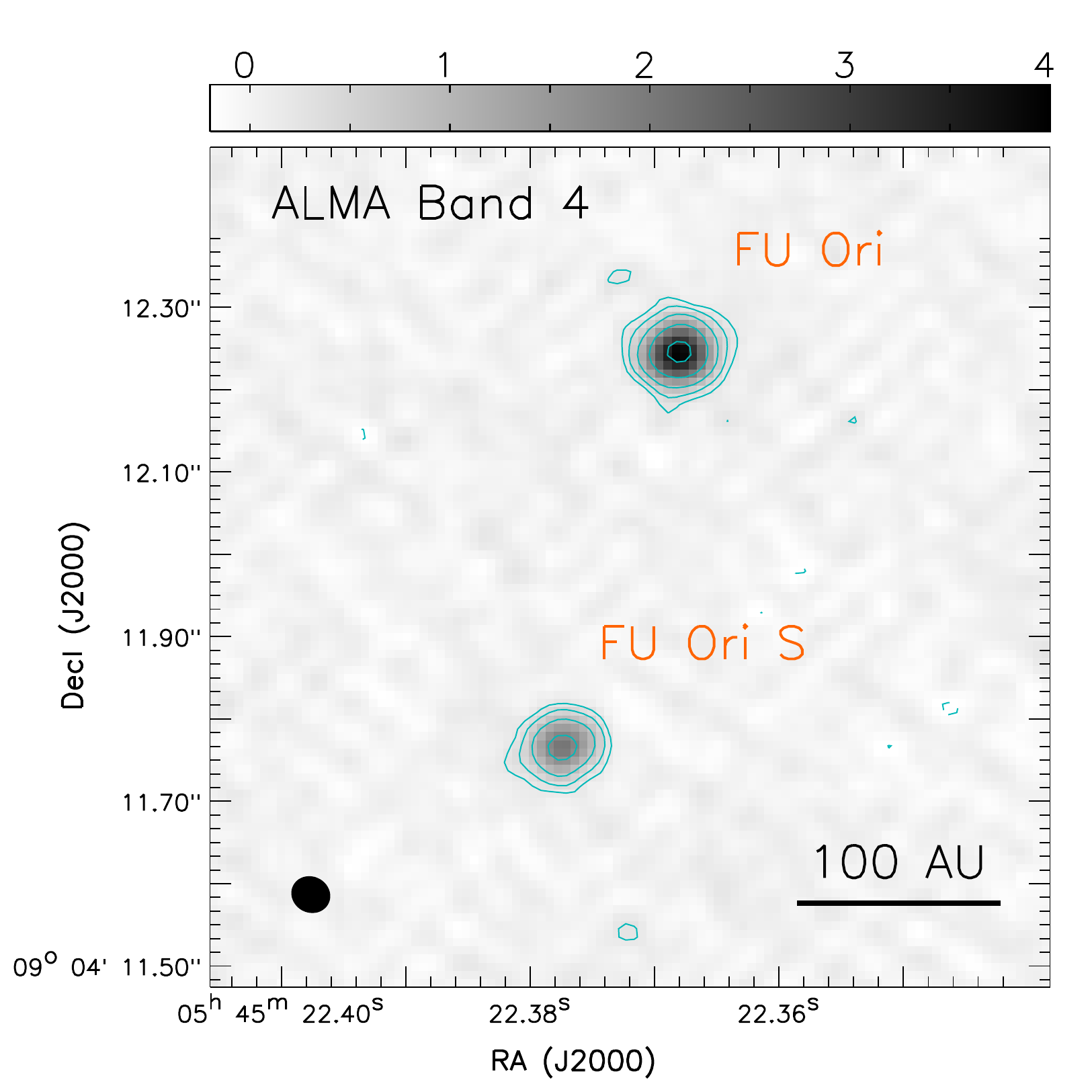} \\
    \end{tabular}
    \caption{Continuum images of FU\,Ori (and S) taken with ALMA at 100 GHz (Band 3, left panel) and 160 GHz (Band 4, right panel), which were generated with 1.875 GHz spectral bandwidth. The synthesized beams of these images are \beam=0$\farcs$082$\times$0$\farcs$075 (P.A.=$-$79$^{\circ}$) and \beam=0$\farcs$047$\times$0$\farcs$043 (P.A.=57$^{\circ}$), respectively. Color bars are in units of mJy\,beam$^{-1}$. Contours in the left and right panels are 0.21 $\mu$Jy\,beam$^{-1}$ ($3\sigma$) $\times$[$-$1, 1, 2, 4, 8] and 0.22 $\mu$Jy\,beam$^{-1}$ ($3\sigma$) $\times$[$-$1, 1, 2, 4, 8, 16], respectively.}
    \label{fig:almaimages}
\end{figure*}

\begin{figure*}
    \hspace{0.1cm}
    \begin{tabular}{ p{3.2cm} p{14cm} }
      \includegraphics[scale=0.4]{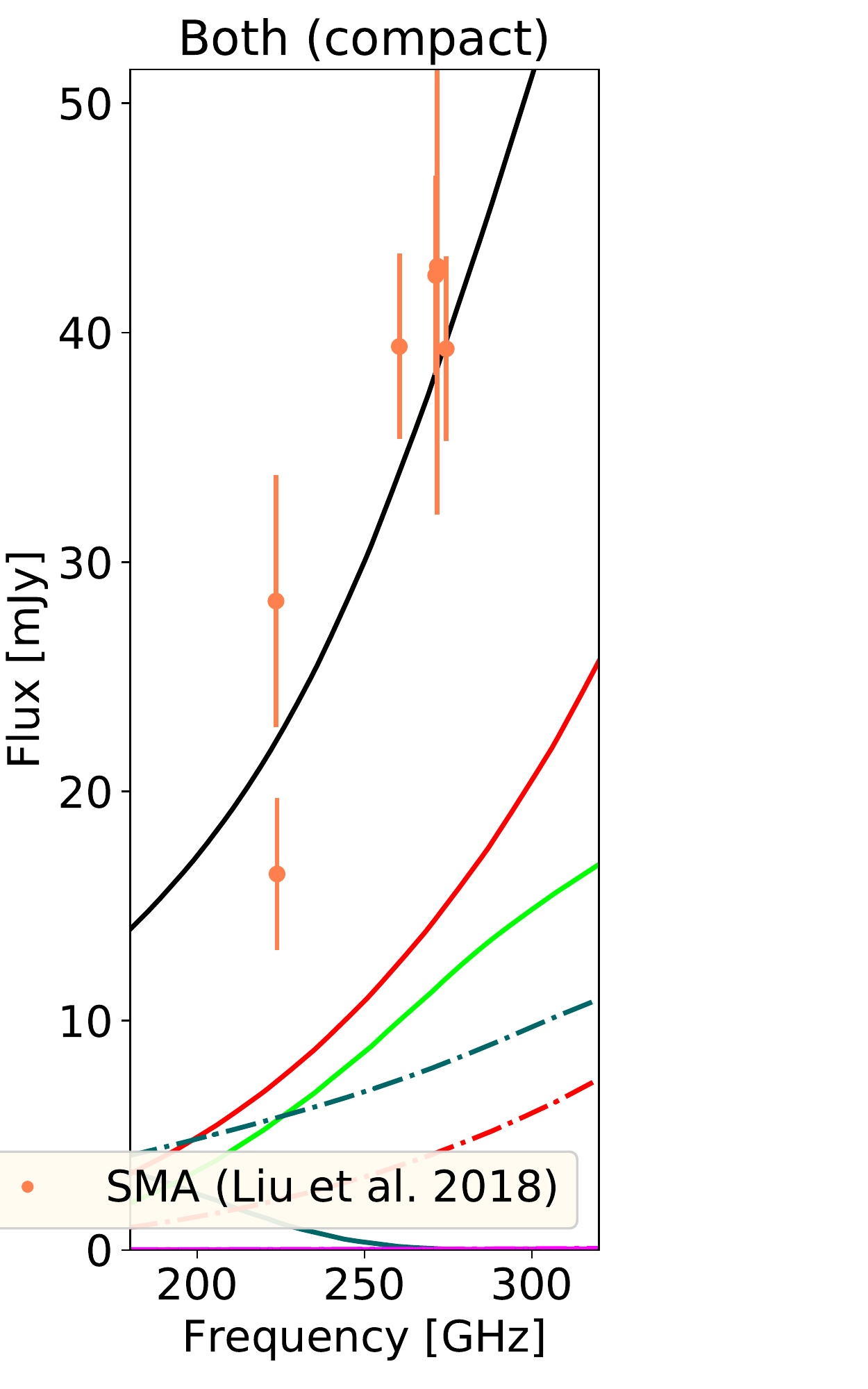} &
      \includegraphics[scale=0.4]{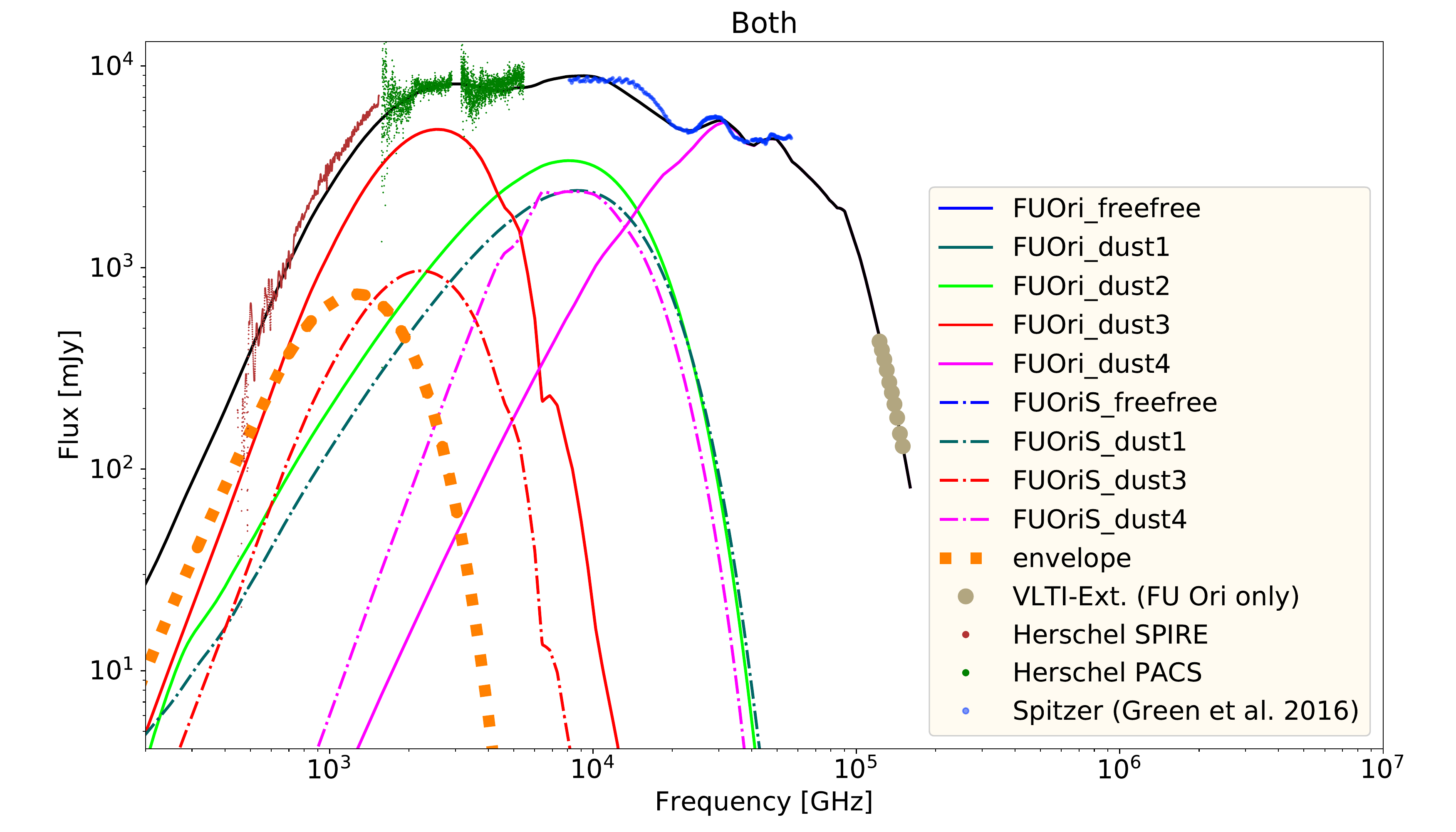} \\
    \end{tabular}
    \caption{Combined fluxes (dots) of FU\,Ori and FU\,Ori\,S taken with the SMA (left panel; \citealt{Liu2018A&A}) and the {\it Herschel} and {\it Spitzer} space observatories (right panel; \citealt{Green2006ApJ,Green2013ApJ,Green2016ApJ}), and the fluxes of the au scales structures around FU\,Ori taken with the VLTI/GRAVITY (i.e., the "Extended" column of Table \ref{tab_flux}). Black lines show our model of the combined fluxes of these two sources.
    Lines with other colors are the fluxes of individual dust or free-free emission components in our model (see also Figure \ref{fig:sedresolved}; c.f., Table \ref{tab:components}).
    We assumed that the envelope component was only detectable from {\it Herschel} and was resolved out by any of our interferometric observations.
    Model components which are labeled but cannot be found in the right panel are due to that their fluxes are below the plotted range.
    }
    \label{fig:sedunresolved}
\end{figure*}

\begin{figure*}
    % \vspace{-6cm}
    \hspace{-2cm}
    \begin{tabular}{ p{10cm} }         
      \vspace{0cm}\includegraphics[width=20cm]{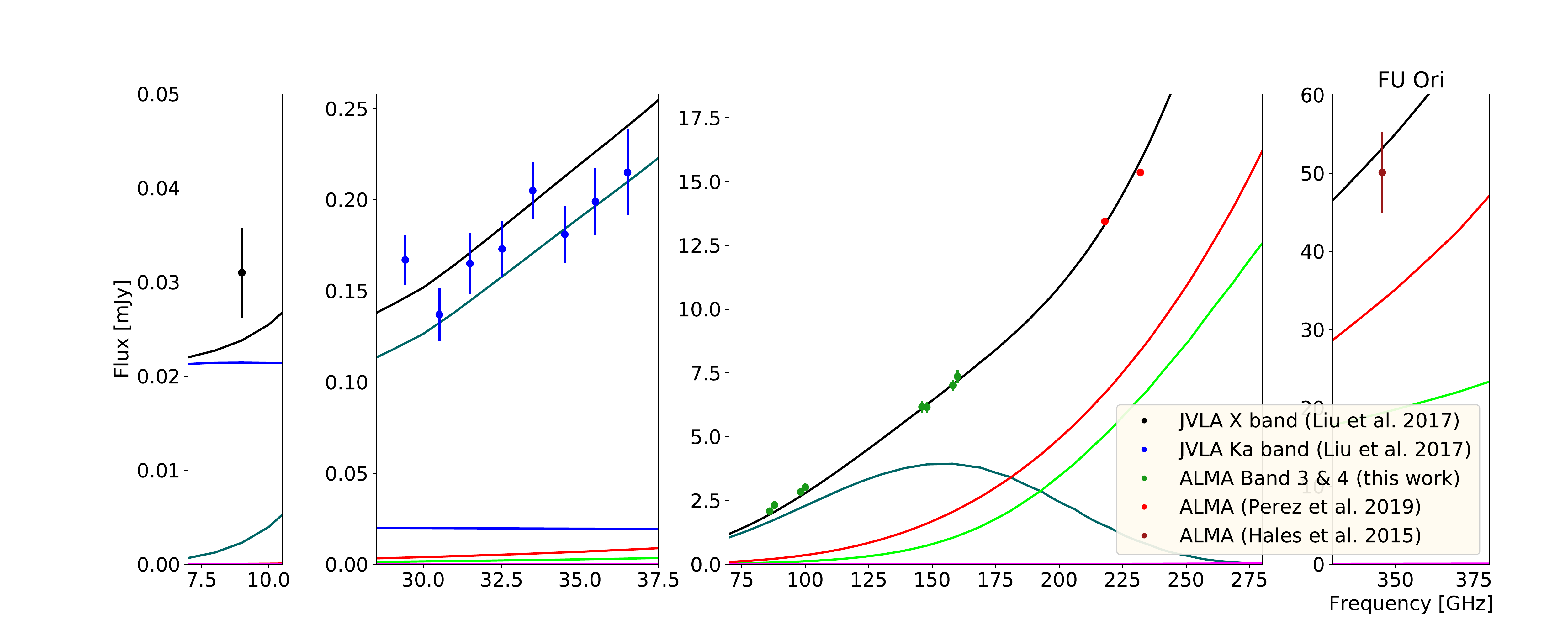} \\
      \vspace{-0.2cm}\includegraphics[width=20cm]{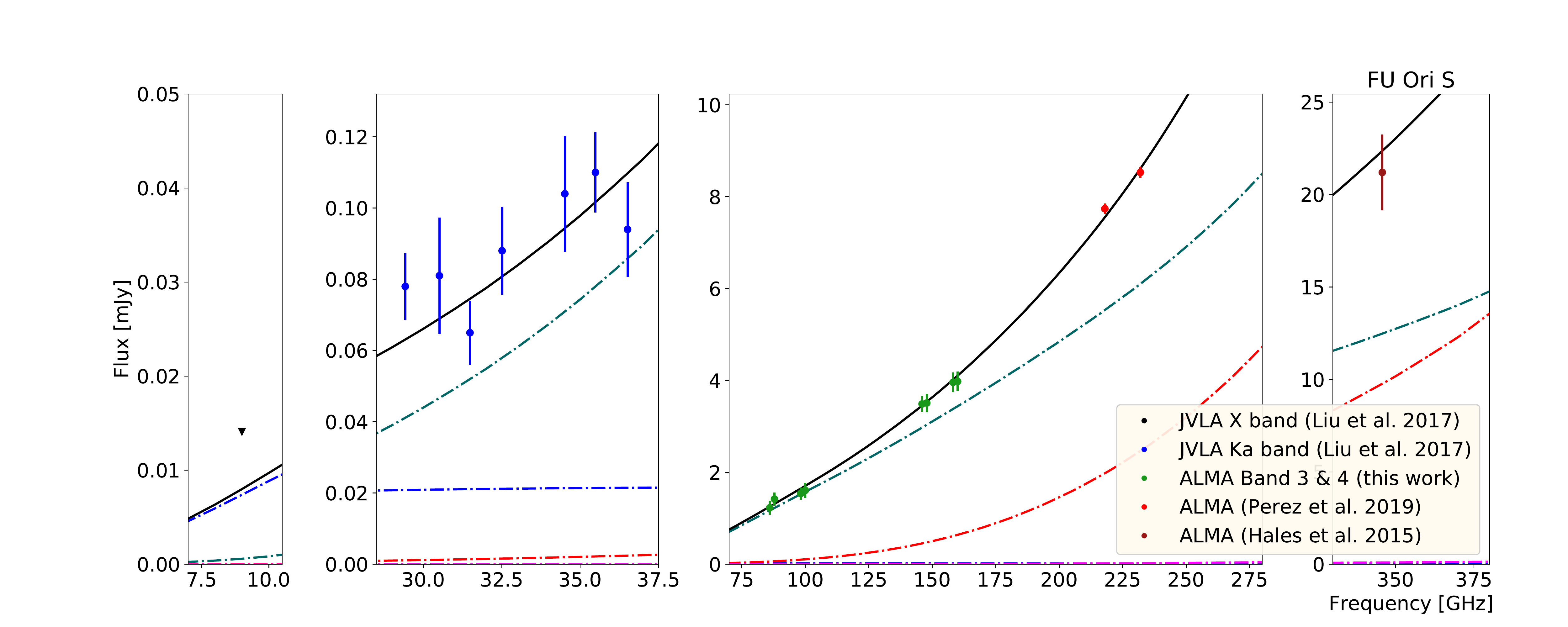} \\
    \end{tabular}
    \vspace{0.5cm}
    \caption{Resolved fluxes of FU\,Ori and FU\,Ori\,S taken with the JVLA \citep{Liu2017A&A} at X band (8-10 GHz) and Ka band (29-37 GHz), and with the ALMA at Band 3 (86-100 GHz), Band 4 (146-160 GHz), Band 6 ($\sim$225 GHz; P\'erez et al. submitted) and Band 7 ($\sim$346 GHz; quoted from \citealt{Hales2015ApJ}). 
    Throughout this paper we assumed a nominal 10\% error for the 346 GHz fluxes of FU\,Ori and FU\,Ori\,S since they were not clearly separated in the previous ALMA image due to the limited angular resolution.
    The upside down triangle shows the 3$\sigma$ upper limit for FU\,Ori\,S at 9 GHz.
    Colored lines show fluxes of our model for each of these two resolved sources (c.f., Table \ref{tab:components}; for the labels see Figure \ref{fig:sedunresolved}). For both sources, blue lines show the free-free emission component; cyan lines show the dense and hot inner disks of a few au scales; red lines show the outer disks on few tens of au scales; the light green line shows a spatially compact dust component which is enclosing the hot inner disk of FU\,Ori and has a lower dust temperature than that of the hot inner disk.}
    % \vspace{-14cm}
    \label{fig:sedresolved}
\end{figure*}

\begin{deluxetable*}{cccccc}
  \tablecaption{Models for the VLTI/GRAVITY data\label{tab_models}}
  \tablehead{
  \colhead{\#} & \colhead{VLTI-compact} & \colhead{VLTI-extended} & \colhead{$\Delta x_e$} & \colhead{$\Delta y_e$} & \colhead{$\chi_r^2$} 
  }
  \startdata
            & uniform disk & 2D Gaussian & & & \\
            & diameter ($mas$) & FWHM ($mas$) & ($mas$) to East & ($mas$) to North & \\
            \noalign{\smallskip}
            \hline
			0& 0.0 & $\infty$ & 0 & 0 & 3.1\\
			1& 0.0 & $4.76\pm0.04$ & 0 & 0 & 1.5 \\
			2& $1.14\pm0.01$ & $\infty$ & 0 & 0 & 1.25 \\
			3& $1.06\pm0.01$ & $8.4\pm0.2$ & 0 & 0 & 1.18 \\
			{\bf 4}& $1.06\pm0.01$ & $7.9\pm0.1$ & $0.6\pm0.1$ & $1.0\pm0.1$ & 1.03 \\
  \enddata
  \tablecomments{
  $\Delta x_e$, $\Delta y_e$, and $\chi_r^2$ are the horizontal and vertical offsets of the VLTI-extended component with respect to the phase referencing center, and the chi-square of the fittings, respectively.}
\end{deluxetable*}

\subsection{ALMA data}\label{sub:almaresult}

Figure \ref{fig:almaimages} shows the ALMA 100 GHz (Band 3) and 160 GHz (Band 4) images.
These ALMA observations detected FU\,Ori and FU\,Ori\,S \citep{Reipurth2004,Wang2004} at high significances. 
However, they only marginally spatially resolved the structures.
Their fluxes determined by fitting two-dimensional Gaussians are summarized in Table \ref{tab:almaflux}.
By quoting the previous JVLA observations at 29-37 GHz \citep{Liu2017A&A} and the ALMA observations at $\sim$225 GHz (P\'erez et al. submitted) and at $\sim$346 GHz \citep{Hales2015ApJ}, the derived (sub)millimeter spectral indices ($\alpha$) at various frequency ranges are summarized in Table \ref{tab:spectralindex}.
Figures \ref{fig:sedunresolved} and \ref{fig:sedresolved} summarize the spectral energy distributions (SEDs) of these two protostars at wavelengths from 2 $\mu$m to 33 mm (9~-~1.5$\times$10$^{5}$ GHz).

Both FU\,Ori and FU\,Ori\,S show spectral index values lower than 2.0\footnote{ The spectral index $\alpha$ was measured assuming that the flux $F_{\nu}$ around a reference  frequency $\nu_{0}$ can be expressed as $F_{\nu}=F_{0}(\nu/\nu_{0})^{\alpha}$} at 29-37 GHz \citep[c.f.,][]{Liu2017A&A}; the averaged $\alpha$ are $\sim$2.5 over the frequency range of 29-100 GHz; the averaged $\alpha$ is approximately 2.0 from 100 to 232 GHz, and are higher than 2.0 at higher frequencies.
We require multiple emission components with distinct physical properties to fit the complex submillimeter spectral slopes in the observed SEDs.
Our detailed SED models for all data presented in Figures \ref{fig:sedunresolved} and \ref{fig:sedresolved} are described in Section \ref{sec:discussion}.

\section{Discussion}\label{sec:discussion}
In Section \ref{sub:gravitymodel} we introduce a simple geometric model to interpret the VLTI/GRAVITY observations. In addition, we have generated simple radiative transfer models to interpret the SEDs of FU\,Ori and FU\,Ori\,S.
In Section \ref{sub:individual} we introduce how we produced the spectra for individual dust or free-free (i.e., from ionized gas) emission components in our radiative transfer model.
In Section \ref{sub:integrated}, we introduce how we integrate each of the emission components to the abstracted geometric models to reproduce the integrated SEDs, and how we optimized the model free parameters using the Markov chain Monte Carlo (MCMC) method.
We discuss the physical implications of our models in Section \ref{sub:implication}.

\subsection{Interpreting VLTI/GRAVITY data}\label{sub:gravitymodel}
The fact that the observed squared visibilities in the VLTI/GRAVITY data vary with wavelength (Figure \ref{fig_data}, right panel) implies that the flux ratio of the VLTI-compact and the VLTI-extended components has a wavelength dependence. 

We can quantify this dependence by fitting the data. For simplicity, we assumed that the VLTI-compact and the VLTI-extended components have constant sizes over the wavelength range covered by the VLTI/GRAVITY observations.
In addition, we assumed that the VLTI-compact component is a uniform disk, while the VLTI-extended component is a two dimensional Gaussian of which the aspect ratio is $\sim$1.
We then performed chi-squared fits to determine the sizes of the two components, and to determine the flux ratios as a linear interpolation between eight equally spaced wavelengths between 2.0 and 2.45 $\mu$m (R$\sim$40). 
We tried various combinations of sizes, allowing the two components to be unresolved, partially resolved, or fully resolved. 
Given that the observed closure phases are less than 2 degrees, to avoid over-fitting, the two components were concentric in most trials. However, in one of the trials, we also explored how much their centers can deviate.

\begin{figure*}
  \hspace{-0.3cm}
  \includegraphics[width=19cm]{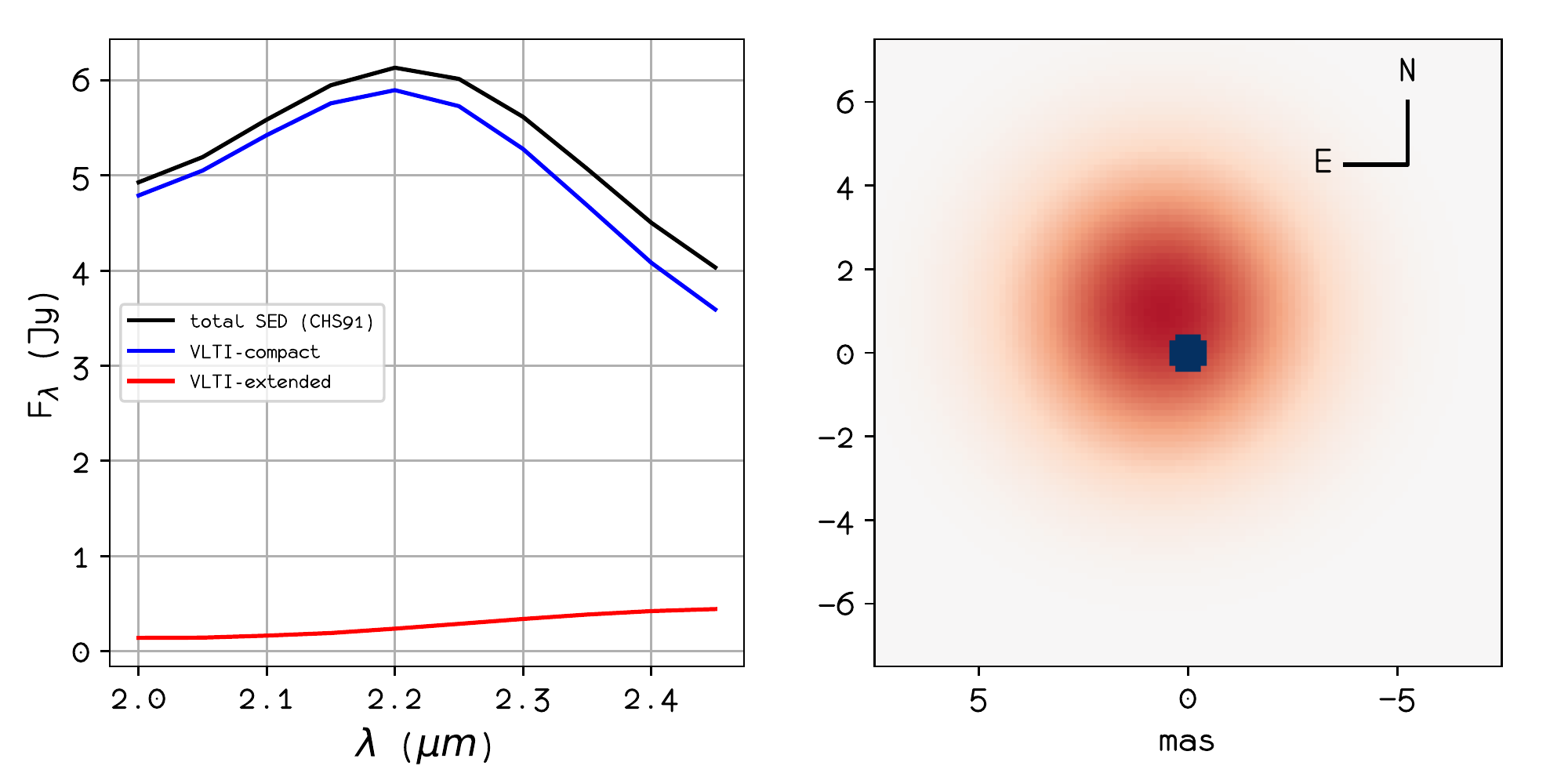}
  \caption{
  Results of model fits to the VLTI/GRAVITY data.
  The left panel shows spectral energy distributions (SEDs) of the unresolved (VLTI-compact) and resolved  (VLTI-extended) components derived from our VLTI/GRAVITY model (Table \ref{tab_models}; see also Table \ref{tab_flux}), and observed total SED quoted from \cite{Calvet1991ApJ} and \cite{Mould1978ApJ}. The right panel plots model \#4 from Table \ref{tab_models}, where the VLTI-compact and VLTI-extended components are shown in dark blue and red, respectively. The horizontal and vertical axes of the right panels are in units of milliarcseconds.}
    \hspace{0.3cm}
  \label{fig_flux}%
\end{figure*}

Our best-fit geometric models are summarized in Table \ref{tab_models}.
The model in best agreement with data is a uniform disk of $\sim$1 mas in diameter, and a two dimensional Gaussian with FWHM$\sim$8 mas (i.e., solid angle $\sim$1.7$\times$10$^{-15}$ sr). 
The fit is further improved if the VLTI-extended component is slightly offset to the Northeast by $\sim$1.2 mas (Figure \ref{fig_flux}, right panel).

Assuming that the VLTI/GRAVITY detections arose predominantly from the circumstellar disk, this spatial offset can be interpreted either as a disk that is geometrically thick (e.g., flared) and is slightly inclined \citep[e.g., Figure 5 of][]{Zhu2008ApJ}, or as a disk that includes substructures or companions \citep[e.g.,][]{Malbet2005}.
Using the flux ratios from Table \ref{tab_flux}, we were able to reproduce the observed slight closure phase signal which increases with wavelength, and the wavelength-dependent variations of the squared visibilities (Figure \ref{fig_data}).
By implementing an absolute flux scaling, the spectral shape of the VLTI-compact component in our best fit model (Figure \ref{fig_flux}) appears fully consistent with the viscous accretion disk model of \citet{Calvet1991ApJ}, which assumed a $\sim$10$^{-4}$ $M_{\odot}$\,yr$^{-1}$ mass accretion rate.
The triangular shape of the spectrum presented in Figure \ref{fig_flux}, following the framework of \citet{Calvet1991ApJ}, is due to the absorption of the water band and the first-overtone vibration-rotation CO band against the bright continuum emission from the viscously heated mid-plane. \citet{Calvet1991ApJ} suggested that these absorption features are predominantly produced at radii of 0.1-0.3 au (i.e., 0.96$\pm$0.48 mas angular diameter assuming $d\sim$416 pc) around the host protostar, which is consistent with the angular sizes in our model fits.

\begin{deluxetable}{ cccc }

  \tablecaption{Fluxes of the best model for the VLT/GRAVITY data\label{tab_flux}}
  \tablehead{
  \colhead{$\lambda$}  &  \colhead{VLTI-compact} &  \colhead{VLTI-extended}  &  \colhead{Ratio} 
  }
  \startdata
            ($\mu$m) & (Jy) & (Jy) & $\pm0.001$\\
            \noalign{\smallskip}
            \hline
            2.00 & 4.40 & 0.13 & 0.030 \\
            2.05 & 5.34 & 0.15 & 0.029 \\
            2.10 & 5.92 & 0.18 & 0.031 \\
            2.15 & 6.15 & 0.21 & 0.034 \\
            2.20 & 6.00 & 0.24 & 0.040 \\
            2.25 & 5.40 & 0.27 & 0.051 \\
            2.30 & 4.79 & 0.31 & 0.065 \\
            2.35 & 4.20 & 0.35 & 0.083 \\
            2.40 & 3.73 & 0.39 & 0.104 \\
            2.45 & 3.44 & 0.43 & 0.125 \\
  \enddata
\end{deluxetable}

\subsection{Individual emission components}\label{sub:individual}
To evaluate the emission properties of dust, we utilized the DSHARP dust optical constants published in \citet{Birnstield2018ApJ}.
For simplicity, we assumed a constant 170 K water ice sublimation temperature \citep{Pollack1994}.
Therefore, we adopted the default DSHARP optical constants for dust emission sources which are cooler than 170 K, and adopted the ice-free optical constants for those which are warmer.
Given that the physical conditions of the observed sources (in particular, FU\,Ori) may be out of equilibrium in various ways, it is not possible for us to evaluate the detailed form of the grain size distribution function from first principles.
Therefore, when evaluating the size-averaged dust absorption ($\kappa_{\nu}^{\mbox{\scriptsize abs}}$) and effective scattering ($\kappa_{\nu}^{\mbox{\scriptsize sca,eff}}$) opacities, we simply assumed the typical power-law grain size distribution with a power-law index $q=$3.5, the minimum grain size $a_{\mbox{\scriptsize min}}=$10$^{-4}$ mm, and the maximum grain size \amax.
Before considering mutual obscuration, the SEDs of individual dust emission components ($F^{\mbox{\scriptsize dust}}_{\nu}$) were evaluated based on the analytic radiative transfer solutions published in \citet{Birnstield2018ApJ}.
Motivated by the small (or negligible) angular offsets of the unresolved and resolved components in the VLTI/GRAVITY models (Table \ref{tab_models}; see discussion in Section \ref{sub:gravitymodel}), we considered all dust slabs to be approximately face-on.
We note that introducing inclinations of the dust slabs will not change the conclusion from our radiative transfer models qualitatively. However, this would increase the total number of free parameters.

\begin{figure}
    \vspace{-0.2cm}
    \hspace{-0.8cm}
    \begin{tabular}{ p{10cm} }         
      \vspace{0cm}\includegraphics[width=9cm]{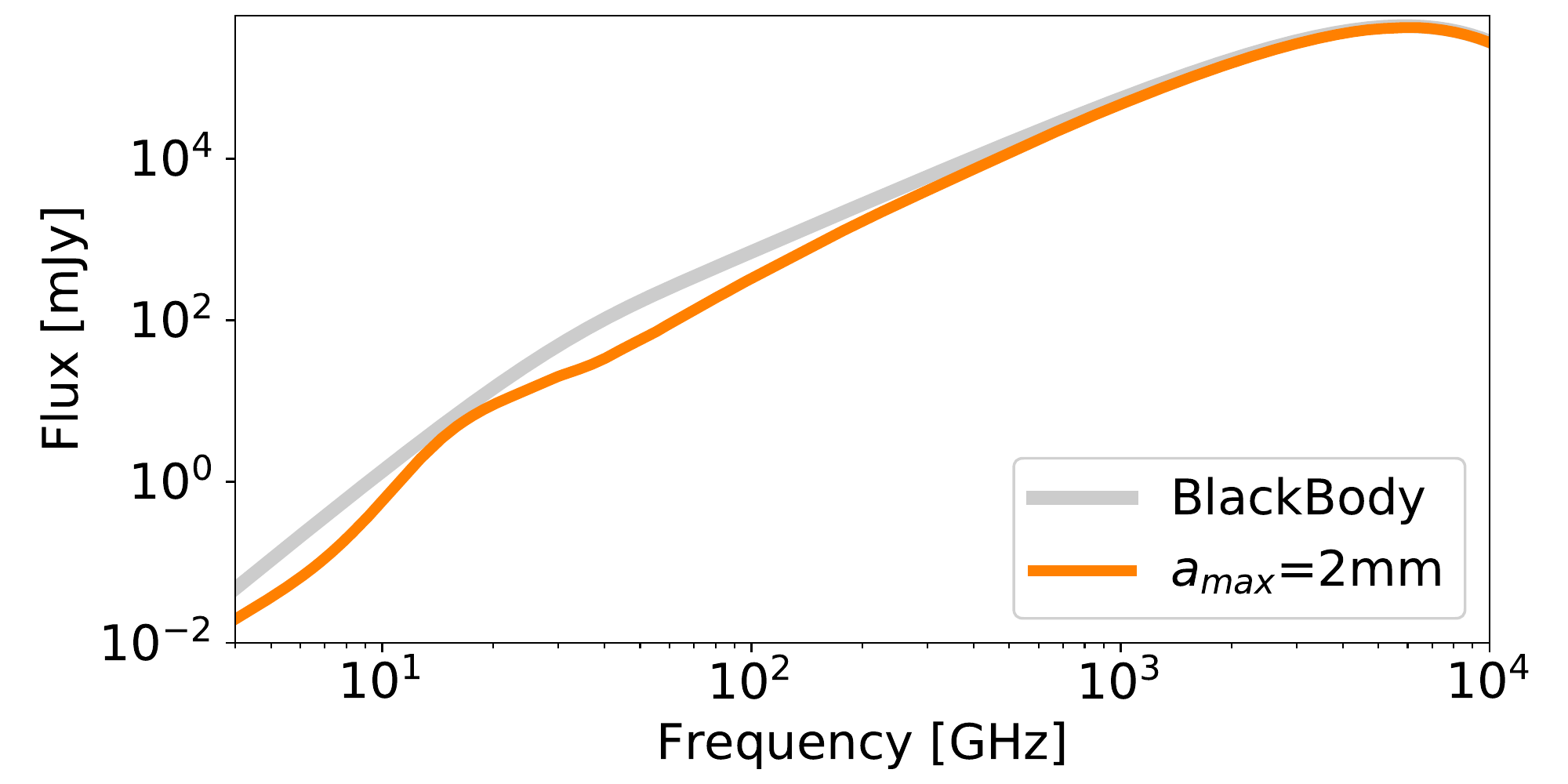} \\
      \vspace{-0.4cm}\includegraphics[width=9cm]{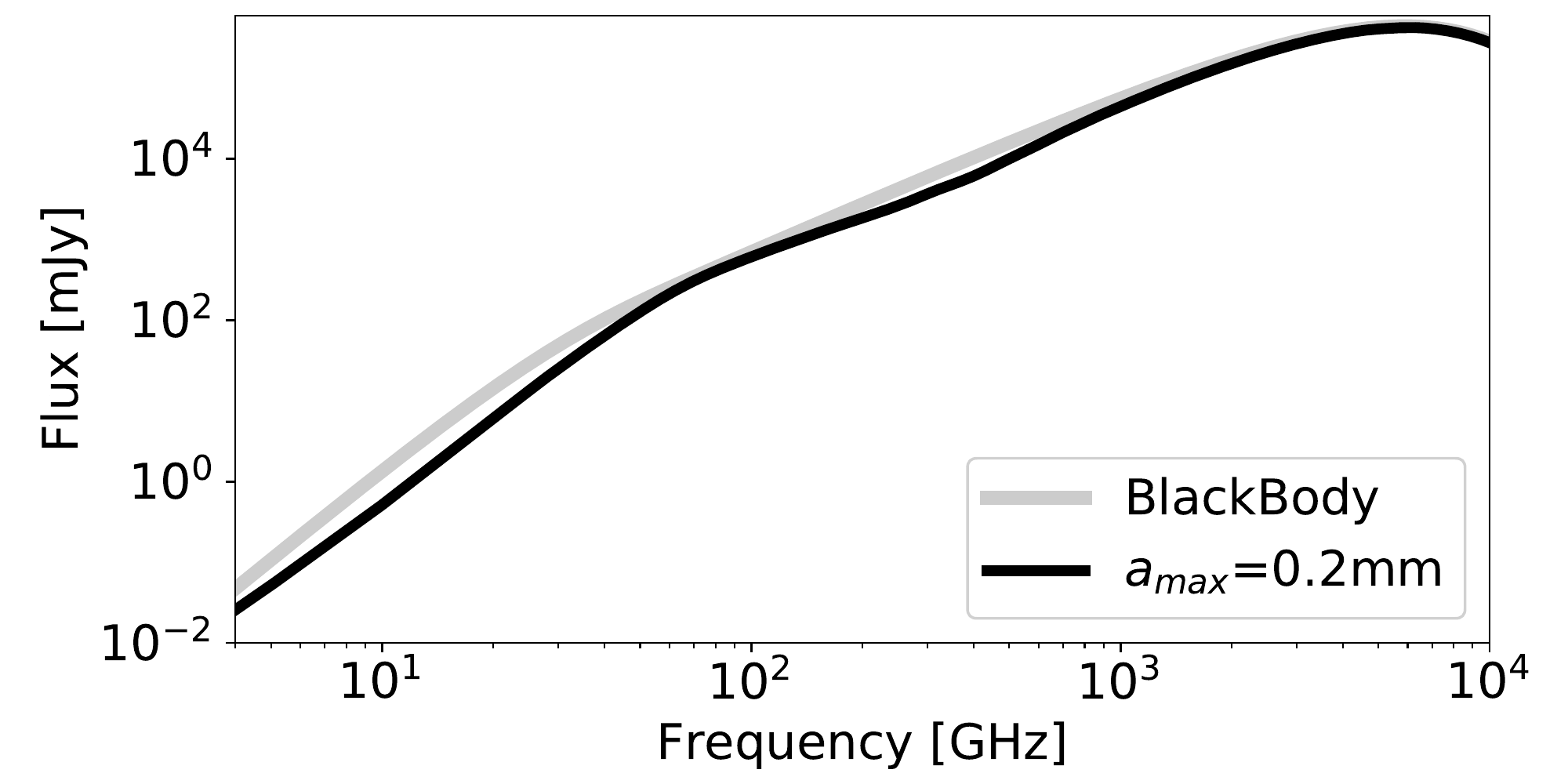} \\
      \vspace{-0.4cm}\includegraphics[width=9cm]{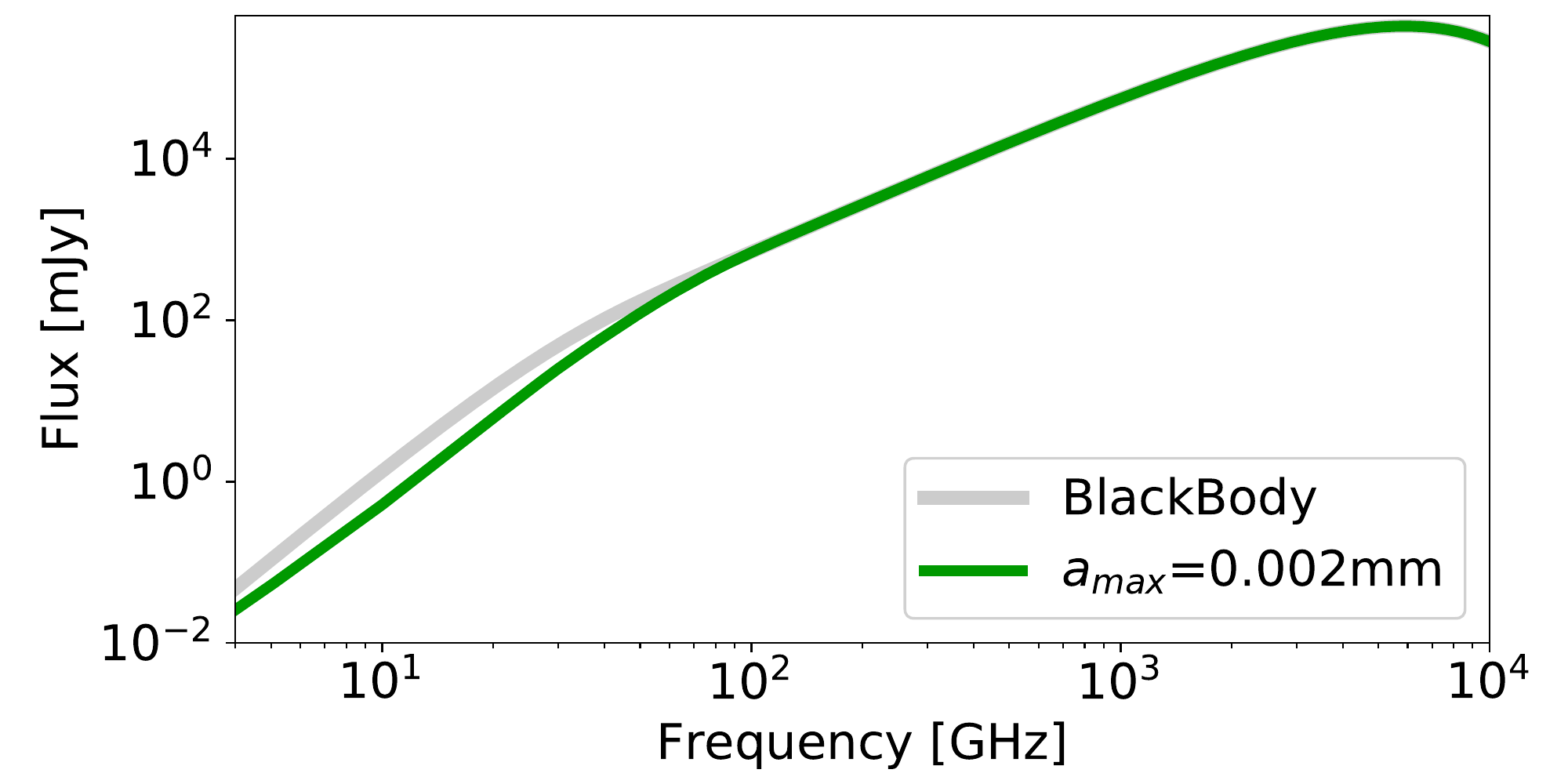} \\
    \end{tabular}
    \vspace{-0.3cm}
    \caption{Spectral energy distributions evaluated for $\Sigma_{\mbox{\scriptsize dust}}=$50 g\,cm$^{-2}$ isothermal (100 K) dust slab of 1 square arcsecond angular size, based on the analytic radiative transfer solution and the dust opacities published in \citet{Birnstield2018ApJ}. Gray dashed line shows the case of black body emission.}
    \label{fig:sedexample}
\end{figure}

Figure \ref{fig:sedexample} shows examples of the SEDs produced for the dust slabs with dust column density of 50 g\,cm$^{-2}$, temperature of 100 K, solid angle 1 square arcsecond, and maximum grain sizes of 0.002 mm (top panel), 0.2 mm (middle panel), and 2 mm (bottom panel).
% At the low frequency ends of these SED examples, the dust slabs are already not very optically thick.
In the low frequency, low optical depth regime, the SEDs deviate from the blackbody emission model (i.e., Planck function) as dust grains cannot emit/absorb efficiently at wavelengths which are much longer than \amax.

In addition, for \amax$=$0.2 mm or 2 mm, the SED deviates from a blackbody curve at higher frequencies. As frequency increases, the spectral indices fall below a blackbody curve, and then become steeper; thus the flux in this frequency regime is below that of a blackbody curve. The dust slabs are optically thick in this frequency regime, and the effects of dust (self-)scattering are not necessarily negligible. We attribute this deviation from blackbody emission in the high frequency regime to the frequency variations of albedo, which was addressed in detail in \citet{Liu2019} and \citet{Zhu2019}.
For example, in Figure \ref{fig:sedexample}, the SED of the \amax$=$2 mm dust slab shows a rather flat spectral index at $\sim$20-50 GHz, which is because the albedo increases with frequency; the spectral index is steepened at $\sim$50-1000 GHz because the albedo decreases with frequency.

Following \citet{Mezger1967} and \citet{Keto2003}, we approximated the optical depth of the free-free emission components $\tau^{\mbox{\scriptsize ff}}_{\nu}$ by
\begin{equation}
\label{eq:tauff}
\begin{split}
\tau_{\nu}^{\mbox{\scriptsize ff}}= & \\
& 8.235\times10^{-2}\left(\frac{T_{\mbox{\scriptsize e}}}{\mbox{K}}\right)^{-1.35}\left(\frac{\nu}{\mbox{GHz}}\right)^{-2.1}\left(\frac{\mbox{EM}}{\mbox{pc\,cm$^{-6}$}}\right), \\
\end{split}
\end{equation}
where EM is the emission measure defined as EM$=$ $\int n_{\mbox{\scriptsize e}}^{2}d\ell$, with $n_{\mbox{\scriptsize e}}$ being the electron number density, and $\ell$ is the linear size scale of the free-free emission component along the line of sight.
Fluxes of individual free-free emission components were evaluated based on
\begin{equation}
  F^{\mbox{\scriptsize ff}}_{\nu} =  \Omega_{\mbox{\scriptsize ff}}(1-e^{-\tau^{\mbox{\scriptsize ff}}_\nu})B_{\nu}(T_{\mbox{\scriptsize e}}),
\end{equation}
where $\Omega_{\mbox{\scriptsize ff}}$ is the solid angle of the free-free emission component, and $B_{\nu}(T)$ is the Planck blackbody function.

\subsection{Abstracted geometric model and integrated SEDs}\label{sub:integrated}

The overall fluxes ($F_{\nu}$) of FU\,Ori and FU\,Ori\,S were determined from the following formulation:

\begin{equation}\label{eqn:multicomponent}
  \begin{split}
    F_{\nu} = & \sum\limits_{i} F_{\nu}^{i} e^{-\sum\limits_{j}\tau^{i,j}_{\nu}}, \\
  \end{split}
\end{equation}
where $F_{\nu}^{i}$ is the flux of the dust or free-free emission component $i$, and $\tau^{i,j}_{\nu}$ is the optical depth of the emission component $j$ to obscure the emission component $i$.
The abstracted geometric information is provided by $\tau^{i,j}_{\nu}$.
% Due to the yet limited angular resolutions, the observations presented in this paper cannot provide constraints on the detailed shapes of the individuals emission components.
We chose this approach instead of fitting analytical solutions of (gaseous) disks because dusty protoplanetary disks are commonly composed of sub-structures (e.g., rings, crescent, etc).
Our SED fitting procedure for the spatially unresolved target sources effectively decomposed them into sub-structures of certain projected areas but without explicitly constraining the shapes.
In this work we considered a simple implementation, such that $\tau^{i,j}_{\nu}=0$ if the emission component $i$ is not obscured by the emission component $j$; otherwise $\tau^{i,j}_{\nu}=\tau^{j}_{\nu}$.

We tried fitting the observed SEDs with the least number of emission components to minimize the total number of free parameters, for various configurations of $\tau^{i,j}$.
the free parameters and the configurations of $\tau^{i,j}$ were varied interactively, informed by the results of previous trials.
Our interactive fits focused on matching the interferometric data.
Nevertheless, we found that once a good fit for the interferometric data was achieved, the infrared spectrum predicted from the model is also very close to the {\it Spitzer} and {\it Herschel} observations.

After we obtained an approximate fit, we used MCMC to simultaneously optimize all free parameters (i.e., all the parameters in Table \ref{tab:components} except the column of overall dust masses).
We assumed flat priors, which permitted each parameter to vary from half of its initial value to two times of the initial value. 
To prevent the MCMC routine from sampling large unlikely portions of parameter space, we provided an additional constraint from the FU\,Ori\,S 9 GHz non-detection (Figure \ref{fig:sedresolved}). We forced the logarithmic likelihood to be negative infinity when the integrated flux of FU\,Ori\,S at 9 GHz is higher than three times the RMS noise of the observations, a condition in the likelihood function to force rejecting such MCMC samples. 
% {\bf JDG: How about this?}
% \textcolor{red}{Baobab: This case is not detected at 9 GHz. Here I tried to say that if we do not consider the non-detection in the fittings, the MCMC walkers will sample numerous solutions where the 9 GHz flux is too much higher than the upper limit constrained by observation. Therefore, I added a condition in the likelihood function to force rejecting such MCMC samples. This is probably one way MCMC is better than other fitting methods. In general it is tricky to properly handle non-detections.}

The MCMC fittings were initialized with 84 walkers with 1500 iterative steps each; in the end, the results from the first 500 steps were discarded.
The {\it Herschel} and {\it Spitzer} data have very good signal-to-noise ratios, such that their contribution to the likelihood largely outweighted the contribution from interferometric data.
To avoid overfitting the {\it Herschel} and {\it Spitzer} data without achieving a good fit for the interferometric data, we needed to reduce the weight of {\it Herschel} and {\it Spitzer} data.
This is implemented by artificially assigning the flux errors of the {\it Herschel}/SPIRE, {\it Herschel}/PACS, and {\it Spitzer}/IRS to be 1000, 10, and 1 times the detected fluxes.
We have monitored how the likelihood evolved over the MCMC iterations to make sure that the contribution of the {\it Herschel} and {\it Spitzer} data are on the same order with the rest of the data.
We note that during the steps of MCMC, some dust emission components may switch from being based on the default DSHARP optical constants to being based on the ice-free optical constants (i.e., the walkers "walked" from below to above the 170 K dust temperature).
Because of this mid-routine shift, it is very difficult to implement fitting methods other than MCMC.

We found at least four dust emission components are required to fit the JVLA, ALMA, and VLTI/GRAVITY data points for FU\,Ori (for more discussion see Section \ref{sub:implication}).
Therefore, we also adopted a four emission component fit for FU\,Ori\,S.

In addition, we included an extended common envelope component which is required to fit the far-infrared fluxes detected by the {\it Herschel} space telescope.
The common envelope component has an extended angular scale, such that it is filtered out by all interferometric observations presented in this work.
We note that the envelope component must be included since the previous {\it Herschel} photometric imaging observations have spatially resolved complicated structures on sub-parsec scales, which connect to FU\,Ori and FU\,Ori\,S \citep{Green2013ApJ}.
We used a simplified parametric model for the envelope (Table \ref{tab:components}); detailed modeling of the envelope is beyond the scope of our present study.
 
During the iterations, we found that we can obtain a reasonably good fit to the JVLA and ALMA data of FU\,Ori\,S by including only two dust emission components and a free-free emission component.
We tentatively assign one additional, $\sim$140 K dust emission component to FU\,Ori\,S to better explain the {\it Herschel} or {\it Spitzer} spectra at (5-10)$\times$10$^{3}$ GHz.
Qualitatively, the fact that we need this extra component to explain the mid-far infrared spectra indicates that the dust components in our models are not isothermal.
We considered whether  each dust component in our models should be allowed to have a small (e.g., 10\%-20\%) temperature range, which could yield better fits to the infrared spectra.
However, our ability to measure any temperature variation is fundamentally limited by the wavelength-dependent aperture used to extract the {\it Herschel}/SPIRE spectra.
This is because we applied semi-extended source correction to align the two SPIRE modules \citep{Wu2013,Green2016AJ}.
Residual artifacts from this process can bias our SED fits, although we have mitigated this by artificially lowering the weighting of the {\it Herschel} data.
Nevertheless, we do not consider it to be meaningful to use a further detailed parameterization for dust temperature profiles to improve the fittings to our infrared spectra.

Parameters of our best fit model are summarized in Table \ref{tab:components}.
The SEDs of the individual components after incorporating the effect of obscuration, and the integrated SEDs from all emission components, are presented in Figures \ref{fig:sedunresolved} and \ref{fig:sedresolved}.
% We omit presenting the corner plots from MCMC since there are too many free parameters, but can provide it upon requests.

\subsection{Model parameters and their physical implications}\label{sub:implication}
In Sections \ref{subsub:fuori} and \ref{subsub:fuoriS} we discuss qualitatively the fitting parameters for FU\,Ori and FU\,Ori\,S.
The overall geometric picture and the physical implications are discussed in Section \ref{subsub:compare}.

\subsubsection{FU\,Ori model}\label{subsub:fuori}
Qualitatively, the fact that the spectral index $\alpha$ of FU\,Ori is $\sim$2 at 86-232 GHz and is $\sim$3 at frequencies higher than 232 GHz (Table \ref{tab:spectralindex}) indicates that fluxes at intermediate frequency (e.g., $\sim$150 GHz) are a mix of one emission component with $\alpha>$3 and the other emission source with $\alpha<$2:
The $\alpha>$3 component (hereafter FUOri\_dust3) becomes more prominent at higher frequencies, while the $\alpha<$2 source becomes more prominent at lower frequencies; the observed spectral indices at the intermediate frequencies are weighted averages from these two components.

In order to fit the JVLA data at 29-33 GHz and the ALMA data at 86-160 GHz (Table \ref{tab:almaflux}; Figure \ref{fig:sedresolved}), we need to realize the $\alpha<$2 source by combining at least two dust components: a $\sim$400 K component with high dust column density and \amax$\sim$2 mm (hereafter FUOri\_dust1), obscured by a $\sim$130 K component with modest dust column density and \amax$\sim$0.2 mm (hereafter FUOri\_dust2).
The dust temperature of FUOri\_dust1 is consistent with the high dust brightness temperature observed at $\sim$33 GHz \citep{Liu2017A&A}.
The FUOri\_dust1 component, whether or not it is mixed with some free-free emission, naturally explains the $<$2.0 spectral index at 29-37 GHz, and the $\sim$2.5 spectral index in between 29-100 GHz (Table \ref{tab:spectralindex}), due to the albedo effect introduced in Section \ref{sub:individual} (see also Figure \ref{fig:sedexample}).
Being obscured by FUOri\_dust2 makes the spectral index of FUOri\_dust1 much lower than 2.0 at $\sim$100-150 GHz.
FUOri\_dust2 has \amax$\sim$0.2 mm because this \amax value yields a high albedo at $\sim$200 GHz.
In this case, FUOri\_dust2, which is optically thick at $\sim$200 GHz and has a rather flat spectral distribution at this frequency, can scatter off the emission from FUOri\_dust1 without contributing much of the emission.
This is critical to fit the steeper spectral indices at 232-345 GHz including the optically thin dust component FUOri\_dust3.
If one or both of FUOri\_dust1 and FUOri\_dust2 contributes more emission at $\sim$200 GHz, it becomes impossible to reproduce the steep spectral index observed at 232-345 GHz. 

The \amax of FUOri\_dust3 is not well-constrained by the data presented in this paper.
Consistent with previous reports of near-infrared scattered light \citep[e.g.,][]{Liu2016SciA,Takami2018ApJ}, we presume that the \amax of FUOri\_dust3 is on the order of $\sim$2 $\mu$m.
% For the same reason, we expect the \amax of the envelope component to be around $\sim$2 $\mu$m.
We cannot accurately determine the dust masses of these two components due to the uncertainties of the dust mass opacities.

Finally, by introducing another FUOri\_dust4 component we can simultaneously fit the resolved VLTI-extended component (Table \ref{tab_models}) in the VLTI/GRAVITY data and the higher frequency part of the {\it Spitzer} spectrum (Figure \ref{fig:sedunresolved}).
The dust temperature of FUOri\_dust4 ($\sim$700 K) is higher than that of FUOri\_dust1 ($\sim$400 K), indicating that FUOri\_dust4 is likely the closest component to the host protostar.
The \amax value of FUOri\_dust4 is not constrained by the observations presented in this paper.
The resolved VLTI-compact component (Table \ref{tab_models}) is hotter than the dust sublimation temperature and therefore is not considered in our dust models.
The thermal radiation from the VLTI-compact component may heat the VLTI-extended component (c.f., \citealt{Zhu2007ApJ}).
We note that an excellent fit to the 9 GHz observations with the free-free emission component was not necessary, because that particular measurement was impacted by poorly-characterized delay errors, and is rather uncertain (c.f., \citealt{Liu2017A&A}; P\'erez et al. submitted).
Emission at 9 GHz may also include a non-thermal emission contribution, which we do not have sufficient data to constrain.

\subsubsection{FU\,Ori\,S model}\label{subsub:fuoriS}
The spectral index of FU\,Ori\,S is $\sim$ 2 over a broad frequency range of 86-346 GHz (Table \ref{tab:spectralindex}).
An optically thick dust component (FUOriS\_dust1) with \amax$\sim$0.2 mm can explain the slightly smaller than 2.0 spectral index at $\sim$ 90 GHz.
Mixing FUOriS\_dust1 with an optically thinner dust component (FUOriS\_dust3) and a free-free emission component can better fit the observations at 29-37 GHz and at 346 GHz (Figure \ref{fig:sedresolved}).
To reproduce the resolved SEDs for FU\,Ori\,S, there is no need of assuming mutual obscurations of the emission components since its spectral index at 232-345 GHz is not as steep as that of FU\,Ori (Table \ref{tab:components}).

\subsubsection{Outbursting versus quiescent disks?}\label{subsub:compare}

By assuming a geometrically thin, axisymmetric, Keplerian, hot inner disk around the center of FU\,Ori, \citet{Calvet1991ApJ} argued that the observed CO linewidths at near infrared bands are consistent with an inclination of $\sim$20$^{\circ}$-60$^{\circ}$.
Based on analyzing the squared visibilities from near and mid-infrared interferometric observations, \citet{Malbet2005}, \citet{Zhu2008ApJ} and \citet{Quanz2006ApJ} suggested that the inclination of the disk is $\sim$50$^{\circ}$.
However, being an accretion outburst object, FU\,Ori may not be in equilibrium.
The assumptions of geometrically thin, axisymmetry, and the Keplerian velocity fields all need to be tested by resolved observations.
A great advantage of VLTI/GRAVITY over the previous generations near- or mid-infrared interferometry is that we can anchor the hypothesis of small inclination angle based on the resolved small closure phases (Figure \ref{fig_data}; Section \ref{sub:gravitymodel}).
Thus on the spatial scales of a few au, the morphology and the gas kinematics of the FU\,Ori disk may be more complicated than previously assumed, which can be further resolved by future observations with better {\it uv} coverage.

\begin{figure}
    \centering
    \includegraphics[width=9cm]{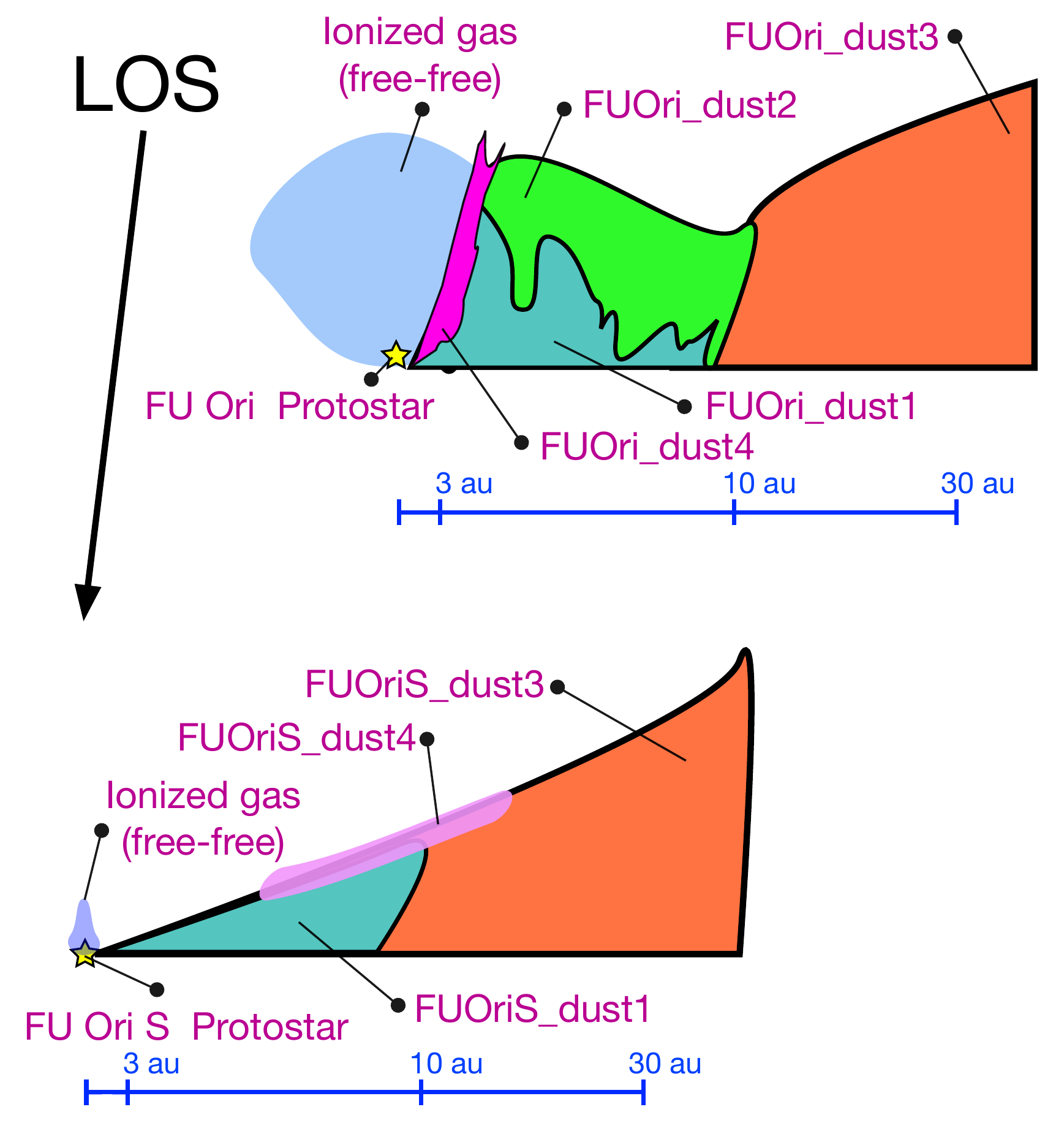}
    \vspace{-0.2cm}
    \caption{Schematic picture of our models for FU\,Ori and FU\,Ori\,S (omitting the envelope component). The colors are chosen only to match the color coding of the SED components in Figure \ref{fig:sedunresolved} and \ref{fig:sedresolved}. 
    The shapes of individual components also do not have strict physical meanings since they were not very well spatially resolved by the observations presented in this manuscript.
    For FU\,Ori, a 1 mas angle corresponds to a spatial scale of 0.416 au.}
    \label{fig:schematic}
\end{figure}

If we assume an approximately face-on projection of FU\,Ori (and FU\,Ori\,S), then the abstracted geometry we introduced during the SED fits (Section \ref{sub:integrated}) follows the picture in Figure \ref{fig:schematic}.
Overall, we interpret the observational data for FU\,Ori as the following: a $>$1000 K hot inner disk at 0.1-0.3 au radii (0.24-0.72 mas) which produces water and CO absorption features at near-infrared bands (unresolved by VLTI/GRAVITY); a $\sim$700 K, not very optically thick dust component with $\sim$3 au radius ($\sim$7 mas; FUOri\_dust4, resolved by VLTI/GRAVITY); a very optically thick and a modestly optically thick dust component with up to $\sim$10 au radii (FUOri\_dust1,2); an optically thin, cooler dust component on tens of au scales (FUOri\_dust3), and some free-free emission.
Assuming that the gas-to-dust mass ratio is $\sim$100, the mass surface density of the component FUOri\_dust1 (Table \ref{tab:components}) is reasonably consistent with the hydrodynamic simulations presented in \citet{Zhu2010ApJ} and \citet{Bae2014ApJ}.
However, the vertical thermal profile of FU\,Ori in its inner 10 au region appears opposite to the typical model of passive disks dominated by radiative heating \citep[e.g., some T Tauri disks;][]{Kama2009,Tapia2017}.
The role of viscous heating in dust thermal dynamics is presently uncertain as it is  difficult to observationally constrain gas volume density and viscosity at the disk mid-plane. 

FU\,Ori\,S can be interpreted as an optically thick dust component with $\sim$10 au radius (FUOriS\_dust1) and an optically thin, cooler dust component on tens of au scales (FUOriS\_dust3), potentially with contributions from free-free emission.

The qualitative difference between the inner $\sim$10 au region of the FU\,Ori and the FU\,Ori\,S disks, in particular the thermal profile, may be related to the thermal and magnetorotational instabilities triggered during the outburst of FU\,Ori. For a physical picture, we refer to Figures 1 and 2 of \citet{Zhu2009ApJ}.
The two dimensional hydrodynamic simulations of \citet{Zhu2009ApJ} have demonstrated that during outburst, within $\lesssim$10 au of the protostar, viscous energy dissipation is sufficient to heat gas at the disk mid-plane to a considerably higher temperature than the gas at the disk surface.
It is not yet very clear to us whether or not this can also explain the vertical dust temperature profile of the FU\,Ori disk.

\begin{figure}
    \vspace{-0.5cm}
    \hspace{-0.8cm}
    \includegraphics[width=10cm]{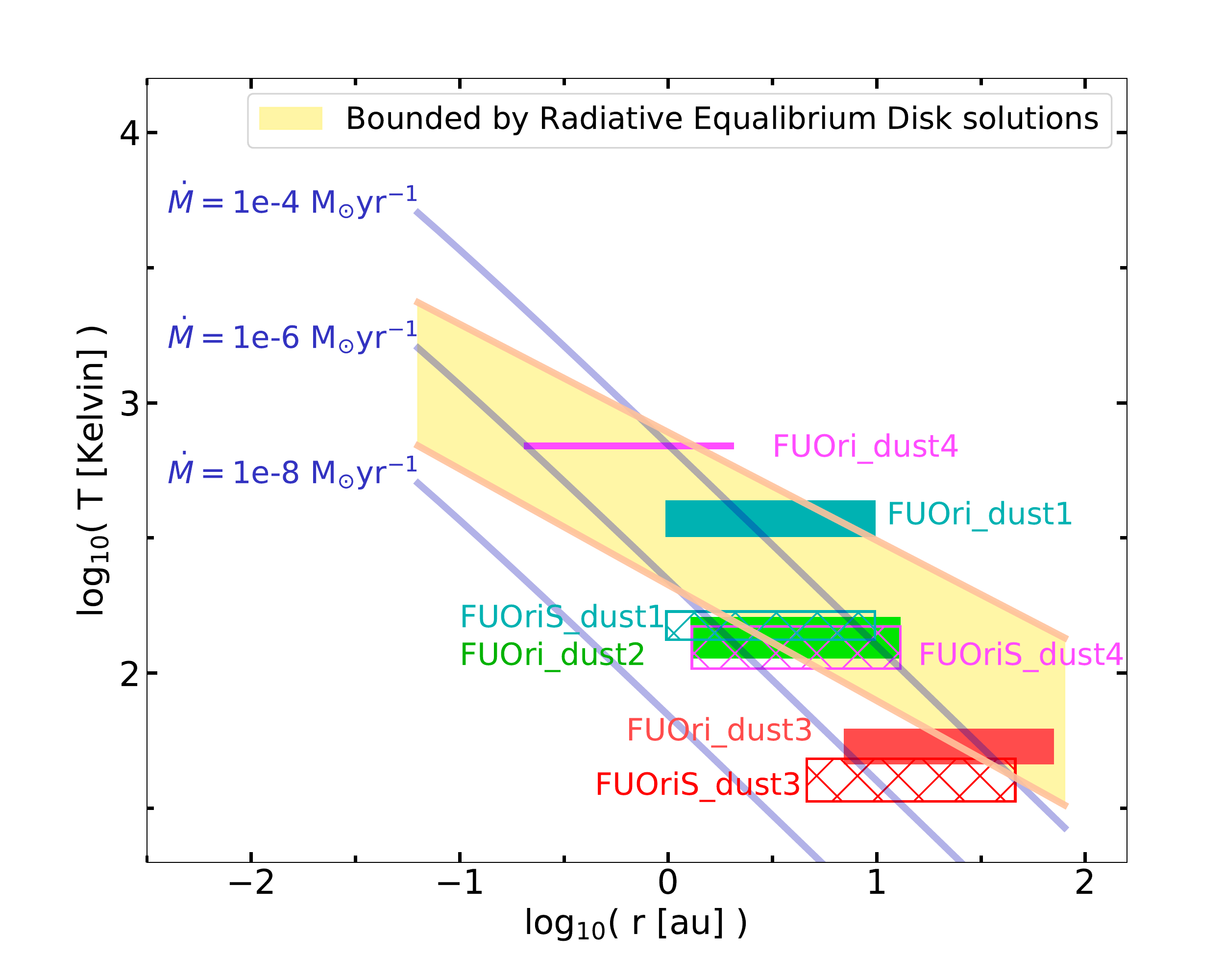}
    \vspace{-0.5cm}
    \caption{A comparison of the fits of dust components with the analytic models of dust temperature profiles due to viscous or radiative heating (see Section \ref{subsub:compare}). The filled and hatched rectangles show the dust components in FU\,Ori and FU\,Ori\,S as listed in Table \ref{tab:components}. The blue lines are the effective temperature profiles of the stead-state viscous disks (stellar mass M$_{*}$=0.5 $M_{\odot}$) with accretion rates $\dot{M}=$10$^{-8}$, 10$^{-6}$, and 10$^{-4}$ $M_{\odot}$\,yr$^{-1}$, in the absence of radiative heating. The yellow filled region is bounded by the surface and interior temperature profiles of a radiative equibrium disk illuminated by a protostar with a 2\,$R_{\odot}$ radius and an effective 4000\,K temperature (c.f., \citealt{Chiang1997ApJ}).
    }
    \label{fig:thermal}
\end{figure}

For an order-of-magnitude estimate, we compare our fits of dust temperatures (Table \ref{tab:components}) with the simplest analytic models of the dust temperature profiles (c.f., \citealt{Chiang1997ApJ} and references therein) in Figure \ref{fig:thermal}.
We caution that many of the underlying assumptions of the simplest analytic models (e.g., axisymmetry, steady or stationary disk, etc.) contradict the observations of the FU\,Ori disk which is likely asymmetric and may be undergoing instabilities over a broad spatial scale.
Such a comparison can serve as a sanity check for whether or not a certain heating mechanism can potentially provide a sufficiently high heating rate to explain the observed dust radiation temperatures.
However, the comparison is not yet sufficient for verifying or strictly falsifying a certain scenario.

To assess how the FU\,Ori disk can be heated due to viscous dissipation, we quoted the effective radiation temperature profile of a steady-state viscous disk $T_{\mbox{\scriptsize viscous}}(r)$ assuming that the disk is very dense and is optically thick such that dust and gas can be thermalized via inelastic collisions, and that there is no radiative heating \citep{Pringle1981}:
\begin{equation}
    T_{\mbox{\scriptsize viscous}}(r) = \left[\frac{3GM_{*}\dot{M}}{8\pi \sigma r^{3}}\left(1 - \sqrt{\frac{R_{*}}{r}}\right)\right]^{\frac{1}{4}},
\end{equation}
where $G$ is the gravitational constant, $M_{*}=$0.5 $M_{\odot}$ is the assumed host protostellar mass, $\dot{M}$ is the mass accretion rate which we assumed to be 10$^{-8}$, 10$^{-6}$, and 10$^{-4}$ $M_{\odot}$\,yr$^{-1}$, $\sigma$ is the Stephen-Boltzmann constant, and $R_{*}$ is the stellar radius which we assumed to be 2\,$R_{\odot}$.
These profiles, which may be regarded as lower limits to the dust temperature in viscous disks, are presented as the blue lines in Figure \ref{fig:thermal}.

To assess how the FU\,Ori\,S disk can be heated due to protostellar irradiation, we scaled the approximate solutions for the surface ($T_{\mbox{\scriptsize s}}(r)$) and interior ($T_{\mbox{\scriptsize i}}(r)$) dust temperature profiles of a radiative equilibrium disk (c.f. Equations 11 and 14a in \citealt{Chiang1997ApJ}) according to the total protostellar luminosity.
The upper and lower bounds of the yellow filled area are shown with respect to $T_{\mbox{\scriptsize s}}(r)$ and $T_{\mbox{\scriptsize i}}(r)$ in Figure \ref{fig:thermal}. 
% \textbf{[JDG: I think I interpreted this sentence correctly, but please check.]}
% \textcolor{red}{Yes!}
We assumed an effective stellar temperature $T_{*}=$4000 K and stellar radius $R_{*}=$2\,$R_{\odot}$, typical for T\,Tauri stars.
We note that due to the Stefan-Boltzmann law the dust temperatures have a weak dependence on the protostellar luminosity.

In Figure \ref{fig:thermal}, we also overplotted our fits of dust components  (c.f., Table \ref{tab:components}).
The inner and outer radii of these dust components were estimated to be 1\% and 100\% of their solid angle, assuming a circular geometry in a face-on projection.
We found that it is plausible to interpret the observed radiation temperature of FUOri\_dust1 based on $T_{\mbox{\scriptsize viscous}}(r)$ given the $\sim$10$^{-4}$ $M_{\odot}$\,yr$^{-1}$ accretion rate of FU\,Ori.
If this is the case, a higher dust temperature at the disk mid-plane than at the surface can be expected, which explains why FUOri\_dust1 has a higher temperature than FUOri\_dust2 (Figure \ref{fig:schematic}).
% In addition, the viscous and radiative heating mechanisms are sufficient to yield the observed dust temperature of the very optically thick component FUOri\_dust1, while the radiative heating alone may be sufficient to explain the dust temperature of the modestly optically thick component FUOri\_dust2.
% Radiative heating alone is sufficient to explain the dust temperature of the optically thick FUOriS\_dust1 component which has a high optical depth.
Moreover, this explains how the 0.1-0.3 au scales hot inner disk with a $10^{-4}$ $M_{\odot}$\,yr$^{-1}$ mass accretion rate (c.f., Section \ref{sub:gravitymodel}) is being replenished by the up to $\sim$10 au scales gas reservoir at a modest rate, such that the hot inner disk neither becomes depleted nor accumulates mass over a short time scales.
This may explain the relatively stable mid-infrared and (sub)millimeter fluxes in the previous monitoring observations \citep{Green2016ApJ,Liu2018A&A}.

The optically thinner components FUOri\_dust3, FUOriS\_dust3,  FUOri\_dust4, and FUOriS\_dust4 are likely dominated by radiative heating.
Radiative heating alone can reasonably explain the observed temperature distributions from FU\,Ori\,S.

The comparisons in Figure \ref{fig:thermal} are uncertain since the accretion rates of FU\,Ori and FU\,Ori\,S are not necessarily constant over all radii. 
In addition, FU\,Ori is unlikely to be in equilibrium, and it is not trivial to accurately estimate the disk scale-height and thus the radiative heating.
Moreover, these comparisons have ignored other mechanical processes which can potentially be important in asymmetric or unstable systems (e.g., shocks, adiabatic compression, etc; \citealt{Dong2016ApJ, Sakai2014Natur}).
More realistic considerations of dust and gas dynamics, grain growth, and dust heating/cooling would provide better comparison. 
We additionally hypothesize that, during the outburst, the inner 0.1-10 au disk may expand significantly in the vertical direction, may be partly thermally ionized, and some dust may be sublimated.
The morphology of the 0.1-10 au disk may also become porous due to accretion and instabilities, allowing dust to be radiatively heated close to the disk mid-plane at a relatively large range of radii.

Finally, why might we have detected millimeter sized \amax from FU\,Ori (i.e., from component FUOri\_dust1) but not from FU\,Ori\,S?
A tentative hypothesis is that at the quiescent stage, dust grains of millimeter or larger sizes may either be radially trapped in regions too small in projected area to be detected by observations (e.g., \citealt{Vorobyov2018,Okuzumi2019}), or areas that are fully obscured due to a combination of very high optical depth and the vertical dust settling.
These mechanisms may be particularly efficient if the inner few au regions are effectively dead zones with negligible ionization fraction during the quiescent stage.
The instabilities during the outburst may help radially and vertically mix dust grains of various sizes, which make the millimeter-sized grains more easily detectable.
That we find tentative evidence of vertical dust settling by comparing the \amax values of FUOri\_dust1 and FUOri\_dust2, may also be because viscous heating is more efficient in heating the vertically settled grown dust from the mid-plane.
This may be further tested by a systematic comparison of the (sub)millimeter and radio spectral indices of the inner disks of outbursting and the quiescent T Tauri sources.

\section{Conclusion}\label{sec:conclusion}
We have analyzed unpublished archival data from the Guaranteed Time Observations of VLTI/GRAVITY at near infrared K-band (2-2.45 $\mu$m) towards the archetypal accretion outburst young stellar object, FU\,Ori.
In addition, we have performed high angular resolution ALMA observations at 86-100 GHz and 146-160 GHz bands, which simultaneously covered FU\,Ori and its companion, FU\,Ori\,S.

The observed small closure phases by VLTI/GRAVITY indicate that the FU\,Ori disk may be approximately face-on.
In addition, by comparing with the squared visibilities resolved by previous generation near and mid-infrared interferometry, we found that the inner few au region of FU\,Ori may not be simply an axisymmetric, Keplerian rotating thin disk.
Instead, it may have a more complicated morphology, which may be related to the instabilities which occurred during the accretion outbursts.
Combined analysis of all existing ALMA, SMA, and JVLA observations along with {\it Spitzer} and {\it Herschel} infrared spectra also points to an unconventional vertical dust thermal profile in the inner $\sim$10 au region of the FU\,Ori disks. This consistently suggests a complicated disk morphology in comparison to a quiescent T Tauri disk.
The observed thermal profiles in the inner $\sim$10 au region may be explained by a viscously heated disk of which the mass inflow rate is $\sim$10$^{-4}$ $M_{\odot}$\,yr$^{-1}$, which can explain how the 0.1-0.3 scales hot inner disk detected from infrared observations is being replenished.

\begin{table*}
  \vspace{0.2cm}
  \caption{Parameters for SED fittings}\label{tab:components}
  \vspace{0.1cm}
  \hspace{-1.3cm}
  \begin{tabular}{ | p{1.3cm} |  p{2.5cm}  p{2.5cm}  p{2.5cm} p{2.5cm} p{2.5cm} p{1.5cm} | }
  \hline\hline
    \multicolumn{7}{| c |}{FU\,Ori} \\\hline
    &  \multicolumn{6}{c | }{Free-free emission} \\
  \hline
          &  $T_{\mbox{\scriptsize e}}$ & EM  & $\Omega_{\mbox{\scriptsize ff}}$ & & Obscured by  & \\
          &  (10$^{3}$ K) & (cm$^{-6}$pc) & (10$^{-14}$ sr) &  & & \\
  \hline
          & 7.0$^{+1.4}_{-1.3}$ & 1.8$^{+4.7}_{-4.7}$$\times$10$^{7}$ & 1.4$^{+0.20}_{-0.21}$ & & none & \\
  \hline\hline
          &  \multicolumn{6}{c | }{Dust components} \\
  \hline
   Comp.  &  $T_{\mbox{\scriptsize dust}}$ & $\Sigma_{\mbox{\scriptsize dust}}$ & \dustang & $a_{\mbox{\scriptsize max}}$ & Obscured by & $M_{\mbox{\scriptsize dust}}$  \\
          &  (K) & (g\,cm$^{-2}$) &  (sr) & (mm) & & ($M_{\oplus}$) \\
  \hline
  1       & 370$^{+63}_{-49}$ & 45$^{+10}_{-12}$ & 4.1$^{+0.64}_{-0.56}$$\times$10$^{-14}$ & 2.4$^{+0.40}_{-0.32}$ & Comp. 2  & 510$^{+210}_{-190}$ \\
  2       & 140$^{+20}_{-26}$ & 0.63$^{+0.12}_{-0.099}$ & 7.1$^{+1.5}_{-1.6}$$\times$10$^{-14}$ & 0.21$^{+0.020}_{-0.019}$ & none  & 12$^{+6.0}_{-3.9}$   \\
  3       & 55$^{+6.8}_{-8.6}$ & 0.13$^{+0.024}_{-0.021}$ & 2.1$^{+0.28}_{-0.28}$$\times$10$^{-12}$ & 0.0017$^{+0.00047}_{-0.00048}$ & envelope  & 75$^{+25}_{-20}$ \\
  4       & 690$^{+18}_{-12}$ & 0.0095$^{+0.0019}_{-0.0018}$ & 1.8$^{+0.34}_{-0.32}$$\times$10$^{-15}$ & 5.2$^{+1.5}_{-1.3}$ & none & 4.7$^{+2.0}_{-1.6}\times$10$^{-3}$ \\
  \hline
  \end{tabular}

  \vspace{0.8cm}
  \hspace{-1.3cm}
  \begin{tabular}{ | p{1.3cm} |  p{2.5cm}  p{2.5cm}  p{2.5cm} p{2.5cm} p{2.5cm} p{1.5cm} | }
  \hline\hline
    \multicolumn{7}{| c |}{FU\,Ori\,S} \\\hline
     &  \multicolumn{6}{c | }{Free-free emission} \\
  \hline
          &  $T_{\mbox{\scriptsize e}}$ & EM  & $\Omega_{\mbox{\scriptsize ff}}$ & & Obscured by  & \\
          &  (10$^{3}$ K) & (cm$^{-6}$pc) & (10$^{-16}$ sr) &  & & \\
  \hline
          & 16$^{+3.7}_{-3.8}$ & 2.1$^{+0.33}_{-0.33}$$\times$10$^{9}$ & 1.9$^{+0.41}_{-0.35}$ & & none & \\
  \hline\hline
          &  \multicolumn{6}{c | }{Dust components} \\
  \hline
   Comp.  &  $T_{\mbox{\scriptsize dust}}$ & $\Sigma_{\mbox{\scriptsize dust}}$ & \dustang & $a_{\mbox{\scriptsize max}}$ & Obscured by  & $M_{\mbox{\scriptsize dust}}$  \\
          &  (K) & (g\,cm$^{-2}$) &  (sr) & (mm) & & ($M_{\oplus}$) \\
  \hline
  1       & 150$^{+19}_{-17}$ & 32$^{+6.4}_{-4.8}$ & 4.1$^{+0.53}_{-0.52}$$\times$10$^{-14}$ & 0.19$^{+0.027}_{-0.025}$ & none & 360$^{+130}_{-90}$ \\
  3       & 41$^{+7.3}_{-7.4}$ & 0.12$^{+0.028}_{-0.023}$ & 9.1$^{+1.9}_{-1.8}$$\times$10$^{-13}$ & 0.0017$^{+0.00047}_{-0.00045}$ & envelope &   30$^{+15}_{-10}$ \\
  4       & 130$^{+19}_{-26}$ & 0.0040$^{+0.0012}_{-0.0011}$ & 7.2$^{+1.4}_{-1.2}$$\times$10$^{-14}$ & 0.0020$^{+0.00047}_{-0.00047}$ & none & 79$^{+41}_{-31}\times$10$^{-3}$  \\
  \hline
  \end{tabular}
  
  \vspace{0.8cm}
  \hspace{-1.3cm}
  \begin{tabular}{ | p{1.3cm} |  p{2.5cm}  p{2.5cm}  p{2.5cm} p{2.5cm} p{2.5cm} p{1.5cm} | }
  \hline\hline
  \multicolumn{7}{| c |}{envelope} \\\hline
         &  \multicolumn{6}{c | }{Dust components} \\
  \hline
   Comp.  &  $T_{\mbox{\scriptsize dust}}$ & $\Sigma_{\mbox{\scriptsize dust}}$ & \dustang & $a_{\mbox{\scriptsize max}}$ & Obscured by & $M_{\mbox{\scriptsize dust}}$ \\
          &  (K) & (10$^{-3}$g\,cm$^{-2}$) &  (10$^{-10}$ sr) & ($\mu$m) & & ($M_{\oplus}$) \\
  \hline
          & 13$^{+3.5}_{-2.7}$ & 5.6$^{+2.3}_{-1.3}$ & 5.7$^{+1.3}_{-1.2}$ & 1.9$^{+0.44}_{-0.44}$ & none &  880$^{+620}_{-350}$ \\
  \hline
  \end{tabular}  

  \vspace{0.2cm}
  \noindent \paragraph{Notes.} $T_{\mbox{\scriptsize e}}$, EM, and $\Omega_{\mbox{\scriptsize ff}}$ are the electron temperature, emission measure, and solid angle of the free-free emission components; $T_{\mbox{\scriptsize dust}}$, $\Sigma_{\mbox{\scriptsize dust}}$, $\Omega_{\mbox{\scriptsize dust}}$,  $a_{\mbox{\scriptsize max}}$, and $M_{\mbox{\scriptsize dust}}$ are the dust temperature, dust mass surface density, solid angle, maximum grain size, and integrated dust mass (in units of earth mass $M_{\oplus}$) of the dust components. The presented values and errors in this table were defined as the 50th and [16th, 84th] percentiles of our MCMC samplers. 1 sr $\sim$4.25$\times$10$^{10}$ square arcsecond.
\end{table*}

\acknowledgments 
% HBL and AM have equal contribution to this manuscript.
This paper is based on data obtained from the ESO Science Archive Facility under request number AMERAND384481.
This paper makes use of the following ALMA data: ADS/JAO.ALMA \#2011.0.00548.S,  \#2016.1.01228.S, and \#2017.1.00388.S. ALMA is a partnership of ESO (representing its member states), NSF (USA) and NINS (Japan), together with NRC (Canada), MOST and ASIAA (Taiwan), and KASI (Republic of Korea), in cooperation with the Republic of Chile. The Joint ALMA Observatory is operated by ESO, AUI/NRAO and NAOJ.
This work has made use of data from the European Space Agency (ESA) mission
{\it Gaia} (\url{https://www.cosmos.esa.int/gaia}), processed by the {\it Gaia} Data Processing and Analysis Consortium (DPAC,
\url{https://www.cosmos.esa.int/web/gaia/dpac/consortium}). Funding for the DPAC has been provided by national institutions, in particular the institutions participating in the {\it Gaia} Multilateral Agreement.
This work is based [in part] on observations made with the Spitzer Space Telescope, which is operated by the Jet Propulsion Laboratory, California Institute of Technology under a contract with NASA.
H.B.L. is supported by the Ministry of Science and
Technology (MoST) of Taiwan (Grant Nos. 108-2112-M-001-002-MY3 and 108-2923-M-001-006-MY3).
E. Vorobyov acknowledges financial support from the Russian Foundation for Basic Research (RFBR), Russian-Taiwanese project \#19-52-52011.
R.G.M. acknowledges support from UNAM-PAPIIT Programme IN104319.
Y.-L. Yang acknowledges the support of University Continuing Graduate Fellowship from The University of Texas at Austin.
The National Radio Astronomy Observatory is a facility of the National Science Foundation operated under cooperative agreement by Associated Universities, Inc.
Y.H. is supported by the Jet Propulsion Laboratory, California Institute of Technology, under a contract with the National Aeronautics and Space Administration.
This project has received funding from the  European Research Council (ERC) under the European Union's Horizon 2020 research and innovation programme under grant agreement No 716155 (SACCRED, PI: \'A. K\'osp\'al).

\facility{ALMA, VLTI/GRAVITY}
\software{CASA \citep{McMullin2007}, Numpy \citep{VanDerWalt2011}, emcee \citep{Foreman-Mackey2013PASP} }

% \clearpage

\bibliographystyle{yahapj}
\bibliography{references}

\end{document}